\documentclass{JFM-FLM_Au}
\usepackage{algorithmic}
\usepackage{float}
\usepackage{placeins}
\usepackage{todonotes}

\newcommand{\RomanNumeralCaps}[1]

\lefttitle{X. Fan, Y. Liu, M. Wang and J.-X. Wang}
\righttitle{Journal of Fluid Mechanics}

\title{Differentiable Hybrid Neural–CFD Modelling of Wall-Bounded Turbulence: Coupled Learning of Subgrid-Scale and Wall Closures}

\author{Xiantao Fan\aff{1}, Yi Liu\aff{1}, Meng Wang\aff{2}, \and Jian-Xun Wang\aff{1}}

\affiliation{\aff{1}Sibley School of Mechanical and Aerospace Engineering, Cornell University, Ithaca, NY, USA
\aff{2}Department of Aerospace and Mechanical Engineering, University of Notre Dame, Notre Dame, IN, USA}

\corresau{Jian-Xun Wang, \email{jw2837@cornell.edu}}

\begin{document}
\maketitle

\begin{abstract}
Wall-modelled large-eddy simulation (WMLES) conventionally treats the subgrid-scale (SGS) closure, wall closure and numerical discretization as independent components, although their effects are coupled through the resolved field. We present a differentiable hybrid neural--CFD framework in which the SGS and wall closures are learned jointly, end-to-end, within a discretized differentiable flow solver using only low-order statistics as training targets. Each closure is a composed neural operator that combines a trainable neural network with a fixed, differentiable layer retaining the physical structure of its conventional counterpart, while preserving solver consistency and physical interpretability. Gradients of the training loss are back-propagated through the coupled solver, allowing both closures to be optimized consistently against the simulated flow rather than fitted offline or in isolation.

We demonstrate the framework, denoted \textit{Hybrid-Joint}, on a zero-pressure-gradient turbulent boundary layer across \textit{a posteriori} configurations spanning $Re_\theta=600$--$6500$, computational domains and mesh resolutions. The model consistently outperforms conventional WMLES baselines, extrapolates without appreciable degradation to more than four times the highest training Reynolds number, and transfers to grids and domains absent from training. It recovers a logarithmic mean-velocity region not imposed by the wall closure and accurately reproduces resolved-scale energy spectra despite spectral information being excluded from the training objective. Ablation studies show that learning either closure alone is insufficient, while joint optimization recovers the full set of statistics, confirming the coupled effects of the SGS closure, wall closure and discretization. Once trained, the closures are reused without retraining across all configurations examined, amortizing the training cost over repeated deployment.
\end{abstract}



\section{Introduction}
\label{sec:intro}

Simulating the unsteady, three-dimensional evolution of wall-bounded turbulence at high Reynolds number remains prohibitively expensive. The total space--time degrees of freedom of a direct numerical simulation (DNS) scale as $N_t N_x N_y N_z \sim Re_{L_h}^{2.91}$, where $N_t$ is the number of time steps, $N_i$ the number of grid points in the $x_i$ direction, and $Re_{L_h}$ a Reynolds number based on the characteristic length of the configuration~\citep{chapman1979computational,yang2021grid}. Wall-resolved large-eddy simulation (WRLES) relaxes this cost only modestly (as $Re_{L_h}^{2.72}$), since resolving the near-wall motions still dominates the expense at high Reynolds number. For outer-loop applications such as design, optimization and uncertainty quantification, the flow must be simulated repeatedly, so the constraint is the cost per simulation rather than the fidelity of any individual run. Reducing this cost requires coarsening the mesh \emph{and} the time step together, so that the space--time degrees of freedom are reduced along every axis.

The difficulty is that coarsening the resolution changes what the simulation can represent: the discretization no longer resolves the full range of turbulent scales, and simply advancing the original governing equations on a coarse grid does not capture the correct dynamics. Two broad strategies have emerged to simulate turbulence in this under-resolved regime. The first replaces the solver altogether with a purely data-driven surrogate, typically an autoregressive neural operator that advances the resolved field from one step to the next~\citep{fukami2019synthetic,guastoni2021convolutional,li2023long,kohl2026benchmarking}. Such surrogates can operate at coarse space-time resolution, but by discarding the governing equations they sacrifice the built-in physical constraints of the solver and tend to compromise stability and generalisability, particularly for three-dimensional high-Reynolds-number turbulence. The second, and in our view more promising, strategy retains the discretized governing equations as a physical backbone and introduces additional modelling/learning only for the scales not resolved by the coarse discretization. This hybrid strategy preserves the structure, conservation properties and boundary conditions of the numerical solver, while relying on closures to account for the unresolved physics.

Within this hybrid strategy, the unresolved contribution can either be learned from data or prescribed from physical reasoning about the unresolved scales. When it is prescribed, the hybrid strategy reduces to the classical framework of wall-modelled large-eddy simulation (WMLES)~\citep{piomelli2008wall,larsson2016large,bose2018wall}, in which two prescribed closures act on the resolved field. In the interior, a subgrid-scale (SGS) model represents the effect of the unresolved small scales, for which a wide range of algebraic and dynamic eddy-viscosity models has been developed~\citep{smagorinsky1963general,germano1991dynamic,vreman2004eddy,bardina1980improved,clark1979evaluation,meneveau2000scale}. Near the wall, where the mesh is too coarse to resolve the small but energetic eddies, a wall model supplies the wall shear stress to approximate the wall boundary conditions for the LES, most used wall model is the stree-balance model that infers the friction velocity from the resolved outer-layer velocity via an assumed instantaneous law of the wall~\citep{piomelli2002wall,piomelli2008wall,bose2018wall}. Casting WMLES in this way, as the physics-based limit of the hybrid strategy rather than as a standalone method, makes clear where its well-known difficulties come from and what a learning-based approach must address.

Conventional WMLES treats these two closures as independent components, yet in practice its accuracy is governed by neither closure alone. A well-known symptom is the log-layer mismatch, a systematic error in the mean velocity profile near the wall. This mismatch can be alleviated by adjusting the matching location~\citep{kawai2012wall}, by modifying the SGS stress model \citep{porte2000scale}, or by filtering the resolved velocity supplied to the wall model \citep{yang2017log}, three routes acting on very different parts of the simulation. That such different interventions all reduce the same error shows that the wall-stress prediction depends on the resolved outer-layer flow, which is itself shaped by the SGS closure and the numerical discretization~\citep{larsson2016large}. The coupling extends to the boundary treatment as well: imposing the wall stress by augmenting the wall eddy viscosity, rather than through a Neumann condition, changes the near-wall statistics~\citep{bae2021effect}. The interdependence can even determine the qualitative flow response: in non-equilibrium flows, whether separation occurs is also governed by the SGS closure rather than the wall model alone~\citep{zhou2024sensitivity,zhou2025effect}. Designing the SGS and wall models separately, as fixed functional forms calibrated in isolation, cannot account for this interdependence~\citep{rezaeiravesh2019systematic}. A second limitation is that these closures are derived at the continuous level and calibrated independently of the numerical solver in which they operate. In the coarse space--time regime, however, the discretization is itself a leading-order source of error: the coarse mesh and large time step introduce numerical dissipation or dispersion that act directly on the resolved field, alongside commutation errors between filtering and differentiation~\citep{huang2026consistency}. Because conventional closures are formulated without reference to a particular discretization, they cannot compensate for these numerical errors, and a model calibrated in the continuum may behave quite differently once embedded in a specific discrete solver. {The SGS closure, wall model and numerical schemes are therefore not separable: they act on the same resolved field and their errors compound.} Conventional WMLES nonetheless designs the two closures independently and in the continuum, leaving their coupling, and their interaction with the discretization, unaddressed.

Machine learning offers a way to relax these fixed functional forms, replacing prescribed closures with models learned from high-fidelity data while retaining the governing equations as a backbone~\citep{wang2017physics,maulik2019subgrid,yang2019predictive}. Most such closures, however, are trained \textit{a priori}: the network is fitted offline to reproduce a target field, such as the SGS stress or the wall shear stress, and only afterwards inserted into the solver. This offline strategy inherits the very limitations it was meant to remove, and adds one of its own. First, fitting each closure separately to its own target reproduces the isolated design of conventional WMLES: the SGS and wall models are learned independently, so their coupling, the very interdependence that governs \textit{a posteriori} accuracy, remains outside the training objective. Second, because the closure never sees the discretized solver during training, good \textit{a priori} accuracy does not guarantee good \textit{a posteriori} behaviour, and models that match their target offline may be inaccurate or unstable once coupled to the solver, where commutation and numerical errors alter their effect~\citep{wu2019reynolds,macart2021embedded,guan2022stable,ling2025numerically,huang2026consistency}. Third, \textit{a priori} training presumes access to the target field, yet reliable pointwise labels for high-order quantities such as the SGS stress are rarely available for high-Reynolds-number wall-bounded turbulence~\citep{mcconkey2021curated,zhang2022ensemble}. These difficulties have motivated a shift towards \textit{a posteriori}, or in-the-loop, training, in which the closure is optimized through the behaviour of the full numerical solver rather than against offline labels~\citep{duraisamy2019turbulence}.

Realizing \textit{a posteriori} training requires gradients of the simulated statistics with respect to the closure parameters, propagated backward through the time-stepping solver. A first route is the adjoint method, which has been used to learn SGS closures for several canonical flows~\citep{macart2021embedded,sirignano2023deep,Kim2023GeneralizableDT}. Deriving and implementing an adjoint solver by hand is laborious, however, and the cost grows with the length of the unsteady rollout. Differentiable programming offers a more general route: by implementing the discretized governing equations in a framework that supports automatic differentiation, the solver itself becomes differentiable, and learnable closures embedded within it can be trained end-to-end against the flow the solver produces. We refer to hybrid models built in this way, in which trainable neural sub-models are coupled to a differentiable numerical solver and optimized through it, as differentiable hybrid neural--CFD modelling \citep{kochkov2021machine,fan2024differentiable,akhare2025hybridndiff,fan2025neural}.

The enabling component is a differentiable CFD platform. A growing number of such CFD solvers have been developed on modern array-programming frameworks (\texttt{JAX} and \texttt{Julia}), including \texttt{JAX-CFD}~\citep{kochkov2021machine}, \texttt{JAX-fluids}~\citep{bezgin2023jax}, \texttt{WaterLily.jl}~\citep{WeymouthFont2025}, \texttt{Diff-FlowFSI}~\citep{fan2026diff}, \texttt{DiFVM}~\citep{du2026difvm}, among others. These platforms leverage a range of modern differentiable-programming techniques to obtain gradients efficiently and stably: reverse-mode automatic differentiation for explicit operations, implicit differentiation, or discrete adjoints, for the implicit components such as the pressure Poisson solve and implicit diffusion, and gradient checkpointing to keep the memory of long space--time rollouts tractable. Together, these techniques allow end-to-end gradients to be propagated stably through three-dimensional, long-horizon simulations, a capability that has been demonstrated for canonical two-dimensional flows~\citep{kochkov2021machine,list2022learned,shankar2025differentiable,fan2025neural} and, more recently, three-dimensional channel flow~\citep{franz2025pict}, though training stability and the admissible rollout length remain delicate~\citep{list2025differentiability}.

Progress along this direction, however, remains partial in ways that matter for wall-bounded turbulence. Many differentiable hybrid-neural modelling studies treat the network as a generic correction to the resolved dynamics rather than as a physically structured closure, and their generalization across flow conditions/regimes has not been established. A more promising way is to make the learnable component itself physically structured, embedding it into the differentiable solver as a constrained closure inspired by conventional WMLES rather than as an unconstrained correction. Such physics-inspired architectural biases were found to be decisive~\citep{shankar2025differentiable}. Work of this kind, however, remains scarce for wall-bounded turbulence. Very recent efforts have so far addressed a single component only: a wall model \citep{zhang2026differentiable} or an SGS closure \citep{Kim2023GeneralizableDT}, but not both. As a result, the coupled effect of the SGS and wall closures, which conventional and \textit{a priori} approaches also fail to capture, has yet to be addressed within a differentiable framework, and the design of objective functions, architectures and training procedures for such stochastic, chaotic flows remains an open question.

In this work, we address this gap by developing a differentiable hybrid neural--CFD framework for simulating unsteady three-dimensional wall-bounded turbulence at coarse space--time resolution, in which the unresolved SGS and wall closures are jointly learned within the discretized flow solver using turbulence statistics as the only training labels. The framework has four main features. First, the two neural closures are structure-constrained rather than free-form, each built on the form of its conventional counterpart, so that physical structure is imposed while only the undetermined part is learned. Second, they are trained jointly and end-to-end, so that their coupling and their interaction with the numerical discretization are optimized consistently against the flow the solver produces. Third, training uses only low-order turbulence statistics, requiring neither instantaneous field matching nor pointwise SGS-stress labels, which makes the approach applicable where only limited statistical data are available. Fourth, the trained model generalizes across Reynolds numbers, computational domains, mesh resolutions and grid distributions, and reproduces the resolved-scale energy spectra accurately on very coarse meshes even though spectral information is never included in the training.

We demonstrate the framework on the canonical zero-pressure-gradient turbulent boundary layer (TBL) across a wide range of Reynolds numbers, for which high-fidelity reference data are available, and leave its extension to non-equilibrium flows, such as adverse-pressure-gradient and separated boundary layers, to future work. The remainder of the paper is organized as follows. Section~\ref{sec:methodology} presents the framework and its components; Section~\ref{sec:results} reports \textit{a posteriori} results for nine test cases; Section~\ref{sec:discussion} analyses what the neural SGS and wall models learn; and Section~\ref{sec:conclusion} summarizes the findings and outlines future directions.

\section{Methodology}
\label{sec:methodology}

\subsection{Problem formulation and closure strategies}
\label{sec:meth_problem}

We consider the incompressible Navier--Stokes equations solved on space--time resolutions far coarser than those required by DNS or WRLES. On such coarse grids the discretization resolves only the large, energy-containing motions, and the effect of the unresolved scales must be reintroduced through modelling. Applying a spatial filter to the incompressible Navier--Stokes equations gives the governing equations for the resolved field,
\begin{equation}
\begin{aligned}
\frac{\partial \overline{u}_i}{\partial t}
+ \frac{\partial}{\partial x_j}\left(\overline{u}_i \overline{u}_j\right)
&= -\frac{1}{\rho}\frac{\partial \overline{p}}{\partial x_i}
+ 2\nu \frac{\partial \overline{S}_{ij}}{\partial x_j}
- \frac{\partial \tau_{ij}}{\partial x_j}, \\
\frac{\partial \overline{u}_i}{\partial x_i} &= 0,
\end{aligned}
\label{eq:filtered_NS}
\end{equation}
where $\overline{u}_i$, $\overline{p}$, $\nu$ and $\overline{S}_{ij}=(\partial\overline{u}_i/\partial x_j+\partial\overline{u}_j/\partial x_i)/2$ denote the filtered velocity, filtered pressure, kinematic viscosity and resolved strain-rate tensor, with $i,j=1,2,3$ the streamwise, wall-normal and spanwise directions. There are two unclosed quantities. The first is the deviatoric SGS stress $\tau_{ij}$, which represents the effect of the unresolved small scales on the resolved field. The second arises at the wall: on a mesh too coarse to resolve the inner layer, a no-slip condition can no longer recover the near-wall momentum transfer, so the wall shear stress $\tau_w$ must be supplied as a model rather than computed from the resolved velocity gradient. Both closures act on, and feed back through, the same resolved field, and their effect is further entangled with the numerical errors introduced by the coarse mesh and large time step, so that the accuracy of the simulation is determined jointly by the two closures and the discrete solver in which they operate.

Two strategies exist for supplying these closures. The conventional strategy prescribes both in fixed functional form, as in standard WMLES. The SGS stress is most commonly represented by an eddy-viscosity model,
\begin{equation}
\tau_{ij} = -2\nu_t\overline{S}_{ij},
\qquad
\nu_t = c\mathcal{F}(\overline{u}_i),
\label{eq:generic_eddyvisc}
\end{equation}
in which the unresolved stress is aligned with the resolved strain rate through a scalar eddy viscosity $\nu_t$, written as a dimensionless model coefficient $c$ times a prescribed functional $\mathcal{F}$ that is built from the resolved velocity field. Eddy-viscosity models differ in the choice of $\mathcal{F}$; the Smagorinsky model, for instance, takes $\mathcal{F}(\overline{u}_i) = |\overline{S}|$~\citep{smagorinsky1963general,lilly1966application}. The wall stress is often supplied by an equilibrium wall model, which assumes local equilibrium in the near-wall region, so that it follows a universal law of the wall,
\begin{equation}
\tau_w = \rho u_\tau^2,
\qquad
\frac{\overline{U}(y_m)}{u_\tau}
= f\left(\frac{y_m u_\tau}{\nu}\right),
\label{eq:generic_wallmodel}
\end{equation}
where $\overline{U}(y_m)$ is the resolved wall-parallel velocity at a matching location $y_m$, $f$ a prescribed profile such as the log law or Spalding's law, and the friction velocity $u_\tau$, hence $\tau_w$, is recovered by inverting this relation. Both closures are thus fixed in advance: the coefficient $c$ and the profile $f$ are calibrated \textit{a priori} or computed dynamically, independently of each other and of the discrete solver on which they run. Such isolated, continuous-level calibration is not ideal in the coarse, coupled regime considered here, since it cannot account for the interdependence of the two closures or for their interaction with the numerical scheme.

The alternative strategy is to learn the closures from data. Instead of prescribing them in fixed functional form, one represents them by ML-based closures that are informed by the conventional forms but learned using high-fidelity data,
\begin{equation}
\tau_{ij} = \mathcal{M}^{\mathrm{sgs}}\left(\overline{u}_{i},\overline{p};\theta_{\mathrm{sgs}}\right),
\qquad
\tau_w = \mathcal{M}^{\mathrm{wall}}\left(\overline{u}_{i},\overline{p};\theta_{\mathrm{wall}}\right),
\label{eq:neural_closures}
\end{equation}
where $\mathcal{M}^{\mathrm{sgs}}$ and $\mathcal{M}^{\mathrm{wall}}$ are neural closures with trainable parameters $\theta_{\mathrm{sgs}}$ and $\theta_{\mathrm{wall}}$. When such closures are embedded in a differentiable solver and optimized through it, they are trained on the flow the discrete solver actually produces, rather than fitted offline to precomputed targets, so that their coupling and their interaction with the discretization are reflected in the training signal itself. This hybrid, solver-consistent formulation is the approach we adopt. Specifically, we treat the closures and the discretized solver as a single differentiable system, using our GPU-native differentiable CFD platform \texttt{Diff-FlowFSI}. In this framework, the SGS and wall closures are represented by neural networks built upon the conventional WMLES forms, and learned jointly within the differentiable solver from turbulence statistics alone, where the gradients can be propagated through the entire time integration. The hybrid model architecture and differentiable training strategies are presented in following sections: \S \ref{sec:meth_ours} and \S \ref{sec:training}.

\subsection{Proposed differentiable hybrid neural--CFD framework}
\label{sec:meth_ours}
\subsubsection{Overview of the hybrid architecture}

\begin{figure}[t]
  \centerline{\includegraphics[width=\textwidth]{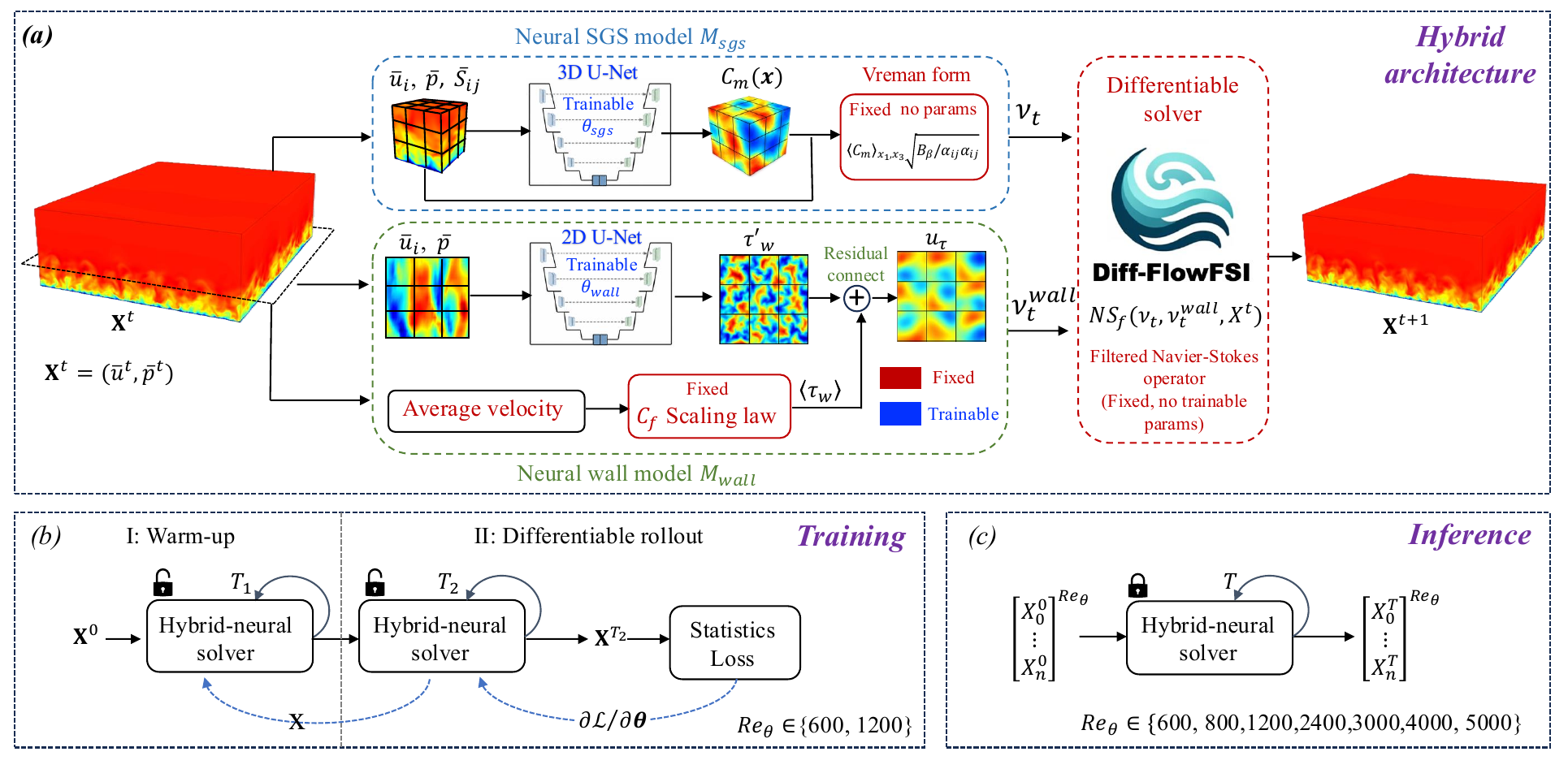}}
\caption{Overview of the proposed differentiable hybrid neural--CFD framework. (a) Hybrid solver architecture over a single time step. The neural SGS closure predicts a spatially varying coefficient that is passed through the fixed Vreman form to obtain the eddy viscosity, while the neural wall closure combines a scaling-law mean wall stress with a learned fluctuating correction. Both closures are coupled with the differentiable Navier--Stokes solver to advance the resolved state. Trainable and fixed components are shown in blue and red, respectively. (b) End-to-end differentiable training with a warm-up window $T_1$ and a gradient-tracking window $T_2$, where the closure parameters are updated from a statistics-based loss. (c) Inference with the trained closures deployed as a conventional WMLES solver over different Reynolds numbers.}
\label{fig:schematic}
\end{figure}

The overall architecture of the proposed hybrid model is shown in figure~\ref{fig:schematic}(a). The framework is built on \texttt{Diff-FlowFSI}, our in-house GPU-native, fully differentiable finite-volume CFD platform for turbulent flow and fluid--structure interaction~\citep{fan2026diff}, which provides the differentiable numerical backbone into which the neural closures are embedded. Over a single time step, the hybrid-neural solver takes the resolved state $\mathbf{X}^t=(\overline{u_i}^t,\overline{p}^t)$, evaluates the two neural closures $\mathcal{M}^{\mathrm{sgs}}$ and $\mathcal{M}^{\mathrm{wall}}$ introduced in equation~\eqref{eq:neural_closures}, and advances the filtered Navier--Stokes equations~\eqref{eq:filtered_NS} to $\mathbf{X}^{t+1}$. The two closures act on the same resolved field and their outputs enter the momentum equation together, so that the solver evolves the flow under their combined effect.

Each closure is constructed as a \emph{composed operator}: a trainable neural network followed by a fixed, differentiable operator that carries the physical structure of conventional closure models. Rather than mapping the resolved field directly to the closure quantity with an unconstrained network, we let the network parameterize only the part that the conventional form leaves undetermined, while the conventional form itself is retained as a fixed layer of the composed hybrid operator. In the SGS closure, a 3-D U-Net outputs a spatially varying coefficient field that is passed to the fixed Vreman form; in the wall closure, a 2-D U-Net supplies a fluctuating wall stress that is added to a mean wall stress estimation from a skin-friction scaling law. In both cases the physical operator is free of trainable parameters, so that the learnable network's influence on each closure is restricted to a scalar coefficient field in the SGS closure and the stress fluctuations in the wall closure, while realizability, near-wall behaviour and invariance are inherited from the retained physical structure.

Because every operation in this sequence, including the U-Nets, the Vreman form, the wall model and the discrete Navier--Stokes update, is differentiable, the single-step map from $\mathbf{X}^t$ to $\mathbf{X}^{t+1}$ is differentiable in the closure parameters $\boldsymbol{\theta}=[\theta_{\mathrm{sgs}},\theta_{\mathrm{wall}}]$. This is what allows the two closures to be trained jointly and consistently with the discrete solver: during training (figure~\ref{fig:schematic}b), the solver is rolled out over a gradient-tracking window and the parameters are updated from a statistics-based loss by backpropagating through the entire sequence of steps, as detailed in \S\ref{sec:training}. Once trained, the closures are frozen and the model is deployed as a conventional WMLES solver (figure~\ref{fig:schematic}c). The specific construction of the two composed closures is presented next.

\subsubsection{Neural SGS closure}
\label{sec:sgs_closure}

The neural SGS closure is a neural operator $\mathcal{M}^{\mathrm{sgs}}$ that maps the instantaneous resolved field to the SGS stress, structured so that its output takes the form of physically admissible eddy viscosity. It consists of a convolutional network backbone followed by a physics-structured output layer, and produces the stress through the eddy-viscosity relation,
\begin{equation}
\tau_{ij} = -2\nu_t\overline{S}_{ij}.
\label{eq:sgs_ansatz}
\end{equation}
The backbone is a three-dimensional convolutional U-Net $\mathcal{N}_{\mathrm{sgs}}$ that maps the instantaneous filtered velocity, pressure and strain-rate tensor, normalized by outer variables, to a spatially varying scalar field,
\begin{equation}
c_m = \mathcal{N}_{\mathrm{sgs}}\left(\overline{u}_i,\overline{p},\overline{S}_{ij};\theta_{\mathrm{sgs}}\right).
\label{eq:cm_net}
\end{equation}
Because the backbone is convolutional, $c_m$ depends non-locally on the surrounding resolved motions rather than on the pointwise velocity gradient alone, and it varies in space and time with the flow. The field is then averaged over the wall-parallel directions, $c = \left\langle c_m \right\rangle_{x_1,x_3}$, leaving a coefficient that varies only in the wall-normal direction. This reduces the learned degrees of freedom, which suppresses spurious streamwise and spanwise variability, and improves robustness during long rollouts.

The output layer maps this learned scalar coefficient field $c$ to an eddy viscosity through $\nu_t = c \mathcal{F}(\overline{u}_i)$ with the Vreman functional~\citep{vreman2004eddy},
\begin{equation}
\mathcal{F}(\overline{u}_i)= \sqrt{\frac{B_\beta}{\alpha_{ij}\alpha_{ij}}},
\qquad
\alpha_{ij} = \frac{\partial \overline{u}_j}{\partial x_i},
\qquad
\beta_{ij} = \Delta_m^2\alpha_{mi}\alpha_{mj},
\label{eq:vreman_form}
\end{equation}
\begin{equation}
B_\beta = \beta_{11}\beta_{22} - \beta_{12}^2
+ \beta_{11}\beta_{33} - \beta_{13}^2
+ \beta_{22}\beta_{33} - \beta_{23}^2.
\label{eq:vreman_Bbeta}
\end{equation}
where $\alpha_{ij}$ is the resolved velocity-gradient tensor, $\alpha_{ij}\alpha_{ij}$ denotes summation over both indices, and $\Delta_m$ is the filter width in each direction, taken as the local grid spacing, since no explicit filter is applied and the filtering is performed implicitly by the mesh. We adopt the Vreman functional as the output structure because of the guarantees it provides for \emph{any} coefficient the network produces: the resulting stress is of eddy-viscosity form and is Galilean and rotationally invariant, and the invariant $B_\beta$ vanishes in laminar and pure-shear regions and near the wall, so that the eddy viscosity switches off there automatically, without the \emph{ad hoc} near-wall damping that a Smagorinsky output layer would require. These properties are enforced by the output layer rather than learned, so the closure does not produce nonphysical SGS stresses regardless of backbone predictions.

The complete closure is thus the composed operator,
\begin{equation}
\tau_{ij}
= \mathcal{M}^{\mathrm{sgs}}\left(\overline{\boldsymbol{u}},\overline{p};\theta_{\mathrm{sgs}}\right)
= -2\left[\left\langle \mathcal{N}_{\mathrm{sgs}}(\overline{u}_i,\overline{p},\overline{S}_{ij};\theta_{\mathrm{sgs}})\right\rangle_{x_1,x_3}
\sqrt{\frac{B_\beta}{\alpha_{ij}\alpha_{ij}}}\right]\overline{S}_{ij}.
\label{eq:sgs_composed}
\end{equation}
Unlike dynamic procedures, which set the coefficient from an instantaneous local-equilibrium assumption and an auxiliary test-filter operation, the backbone learns it \emph{a posteriori}, optimized in the loop against the discrete solver, so that the output prediction reflects the behaviour of the full discretized system rather than a pointwise balance. The trainable backbone thus supplies a flow-dependent, non-local coefficient field, while the Vreman output layer fixes the tensor structure and invariance of the eddy viscosity.

\subsubsection{Neural wall closure}
\label{sec:wall_closure}

The neural wall closure $\mathcal{M}^{\mathrm{wall}}$ supplies the wall shear stress $\tau_w$ as a composed operator: a fixed layer that sets a reference mean stress from a skin-friction scaling law, and a trainable network that corrects this mean and adds the fluctuating component. The two are combined additively,
\begin{equation}
\tau_{w,i} = \langle \tau_{w,i}\rangle + \tau'_{w,i}\left(\overline{\boldsymbol{u}},\overline{p};\theta_{\mathrm{wall}}\right),
\qquad i = 1,3,
\label{eq:wall_decomp}
\end{equation}
where $\langle\tau_{w,i}\rangle$ is the scaling-law mean estimate and $\tau'_{w,i}$ is the network output. This decomposition is motivated by the coarse-resolution regime: at high $Re$ and very coarse near-wall resolution, the instantaneous wall stress is only weakly correlated with the resolved velocity at an off-wall matching point, so a purely local wall-stress relation is unreliable. A scaling law instead ties the mean wall stress to integral boundary-layer quantities and provides the correct $Re$-scaling, which the network need not to relearn; the network then supplies the correction and the fluctuations that the mean-only law omits.

\vspace{0.3em}
\noindent\textbf{Mean estimate from scaling law.} 
For the zero-pressure-gradient boundary layers (ZPG TBL) considered here, the mean streamwise wall stress is set by the empirical skin-friction law of \citet{smits1983low},
\begin{equation}
C_f = \frac{\langle\tau_{w,1}\rangle}{\tfrac{1}{2}\rho U_\infty^2} = 0.024Re_\theta^{-1/4},
\qquad \langle\tau_{w,3}\rangle = 0,
\label{eq:scaling_cf}
\end{equation}
where $\langle\tau_{w,1}\rangle$ and $\langle\tau_{w,3}\rangle$ are the mean streamwise and spanwise wall shear stresses, $U_\infty$ is the free-stream velocity, and $Re_\theta = \theta U_\infty/\nu$ is the Reynolds number based on the momentum thickness $\theta$. This law provides an initial accurate estimate of the mean skin friction over the Reynolds-number range considered here, with an error remaining within approximately 5\% up to $Re_{\theta}=10^{8}$, as reported by \citet{cantwell2021integral}; any comparable correlation, such as the Coles--Fernholz relation~\citep{nagib2023reynolds}, could be used in its place. The momentum-thickness Reynolds number $Re_\theta$ is evaluated online during the simulation. To avoid sensitivity to instantaneous fluctuations, the momentum thickness $\theta$ is computed from a running-time averaged streamwise velocity profile,
\begin{equation}
\widetilde{U}_1^{t+\Delta t}
= \frac{\Delta t}{T}\left\langle \overline{u}_1\right\rangle_{x_3}
+\left(1-\frac{\Delta t}{T}\right)\widetilde{U}_1^{t},
\qquad
\theta = \int_0^\infty \frac{\widetilde{U}_1}{U_\infty}\left(1-\frac{\widetilde{U}_1}{U_\infty}\right)\mathrm{d}x_2.
\label{eq:momentum_thickness}
\end{equation}
where $\left\langle \overline{u}_1\right\rangle_{x_3}$ is the spanwise average of the instantaneous filtered streamwise velocity, $\widetilde{U}_1$ its running time-average, $\delta_{99}$ the $99\%$ boundary-layer thickness, and $T = 10\delta_{99}/U_\infty$ the averaging window~\citep{lund1998generation}, with the resulting $Re_\theta$ is substituted into~\eqref{eq:scaling_cf} to give the reference mean stress.

\vspace{0.3em}
\noindent\textbf{Learned correction and fluctuation.}
The correction is produced by a two-dimensional U-Net $\mathcal{N}_{\mathrm{wall}}$ that takes the instantaneous filtered velocity components and pressure at the matching location, the third off-wall cell centre, and outputs the two wall-parallel stress components $\tau'_{w,i}$ over the wall plane,
\begin{equation}
\tau'_{w,i} = \mathcal{N}_{\mathrm{wall}}\left(\overline{\boldsymbol{u}},\overline{p};\theta_{\mathrm{wall}}\right),
\qquad i = 1,3.
\label{eq:wall_net}
\end{equation}
No zero-mean constraint is imposed on the output; the network therefore both corrects the scaling-law mean and supplies the fluctuations. Because both components are predicted directly, the model does not assume the wall stress to be aligned with the local wall-parallel velocity, as equilibrium wall models do, but can represent a misalignment between the wall stress and the near-wall velocity. Retaining the fluctuating part is important because coherent near-wall stress fluctuations provide unsteady forcing to the resolved flow that a mean-only wall stress cannot reproduce, and that is needed to recover the near-wall momentum transfer at coarse resolution.

\vspace{0.3em}
\noindent\textbf{Imposition of the wall stress. }
Rather than prescribing $\tau_w$ directly as a Neumann condition, its effect is imposed by modifying the wall eddy viscosity while retaining a Dirichlet condition for the velocity~\citep{bae2021effect}. Namely, the wall eddy viscosity is set so that each modelled stress component is recovered at the wall,
\begin{equation}
\nu_t^{\mathrm{wall}}
= \max\left(\frac{\tau_{w,i}}{\rho\partial\overline{u}_i/\partial x_2 + \epsilon} - \nu,\ 0\right),
\qquad i = 1,3,
\label{eq:modified_nu}
\end{equation}
where $\epsilon$ is a small constant preventing division by zero. Clipping the eddy viscosity at zero guarantees $\nu_t^{\mathrm{wall}} \ge 0$ and hence a non-negative total near-wall viscosity, which is required for numerical stability. This bound acts only on the eddy viscosity for imposing the stress at the wall.

The complete wall closure is thus the composed operator,
\begin{equation}
\tau_{w,i} = \mathcal{M}^{\mathrm{wall}}\left(\overline{\boldsymbol{u}},\overline{p};\theta_{\mathrm{wall}}\right)
= \langle\tau_{w,i}\rangle\big(Re_\theta\big) + \mathcal{N}_{\mathrm{wall}}\left(\overline{\boldsymbol{u}},\overline{p};\theta_{\mathrm{wall}}\right), i=1,3.
\label{eq:wall_composed}
\end{equation}
where the scaling law fixes the Reynolds-number scaling of the mean streamwise stress, while the trainable network supplies the flow-dependent correction and the two-component fluctuations, and the stress is imposed through the non-negative wall eddy viscosity.

\subsection{Differentiable training and inference}
\label{sec:training}

The entire hybrid solver is trained as a whole, end-to-end. Namely, the neural SGS and wall closures are not fitted separately or offline, but are optimized jointly against the flow produced by the differentiable solver \texttt{Diff-FlowFSI} in which they are embedded. Because turbulence is chaotic, matching instantaneous fields over a rollout is ill-posed~\citep{fan2025neural}; the training objective is therefore defined in terms of turbulence statistics. At each iteration, the coupled solver is rolled out to accumulate these statistics, a statistics-based loss is evaluated against reference data, and its gradient is back-propagated through the entire rollout to update the neural-closure parameters $\boldsymbol{\theta}=[\theta_{\mathrm{sgs}},\theta_{\mathrm{wall}}]$ in a single optimization. The loss, the rollout and the deployment of the trained model are described below.

\vspace{0.3em}
\noindent\textbf{Statistics-based loss.}
The loss compares the predicted and reference profiles of the mean streamwise velocity, the root-mean-square (r.m.s.) velocity fluctuations and the r.m.s.\ pressure fluctuations, together with the friction velocity. All statistics are expressed in inner (wall) units, denoted by a superscript $+$: velocities are normalized by the friction velocity $u_\tau = \sqrt{\tau_w/\rho}$, with $\tau_w = (\tau_{w,1}^2 + \tau_{w,3}^2)^{1/2}$ the wall-stress magnitude, and the wall-normal coordinate by $y^+ = x_2 u_\tau/\nu$, so that $\overline{u}_i^+ = \overline{u}_i/u_\tau$, $\overline{u}_{i,\mathrm{rms}}^+ = \overline{u}_{i,\mathrm{rms}}/u_\tau$ and $\overline{p}_{\mathrm{rms}}^+ = p_{\mathrm{rms}}/(\rho u_\tau^2)$. Here $u_\tau$ is evaluated from the model-predicted wall stress, while $u_{\tau,\mathrm{GT}}$ denotes the reference value. The loss is defined as
\begin{equation}
\begin{aligned}
\mathcal{L}(\boldsymbol{\theta})
=\;& w_1 \left\| \left\langle \overline{u}_1^{+}\right\rangle - \left\langle \overline{u}_{1,\mathrm{GT}}^{+}\right\rangle \right\|_2
+ \sum_{i=1}^{3} w_{2,i}\left\| \overline{u}_{i,\mathrm{rms}}^{+} - \overline{u}_{i,\mathrm{rms,GT}}^{+}\right\|_2 \\
&+ w_3 \left\| \overline{p}_{\mathrm{rms}}^{+} - \overline{p}_{\mathrm{rms,GT}}^{+}\right\|_2
+ w_4 \left| u_\tau - u_{\tau,\mathrm{GT}}\right|,
\end{aligned}
\label{eq:loss}
\end{equation}
where $\langle\overline{u}_1^{+}\rangle$ is the mean streamwise velocity profile, $\overline{u}_{i,\mathrm{rms}}^{+}$ the r.m.s.\ velocity-fluctuation profiles and $\overline{p}_{\mathrm{rms}}^{+}$ the r.m.s.\ pressure fluctuation, all as functions of the wall-normal coordinate; the subscript GT denotes the reference (ground-truth) data, and $\|\cdot\|_2$ is the $L_2$ norm over the wall-normal profile. Because $y^+$ depends on the model through the predicted $u_\tau$, the predicted and reference profiles are interpolated onto a common $y^+$ grid before the norm is evaluated. The mean and fluctuation profiles constrain the inner-scaled shape of the solution, while the final term anchors the absolute magnitude of $u_\tau$, which inner scaling alone leaves undetermined.
The weights $w_1$, $w_{2,i}$, $w_3$ and $w_4$ balance the terms so that each contributes comparably. Each is set to the inverse magnitude of its term evaluated at the current epoch, $w_k = 1/|\mathcal{L}_k|$, and detached from the computational graph during back-propagation, so that the weighting rescales the terms without contributing to the gradient.  Because the weights are recomputed each epoch, the effective objective is non-stationary, but the terms remain balanced as their magnitudes evolve during training.

\vspace{0.3em}
\noindent\textbf{Differentiable rollout.}
The statistics in equation~\eqref{eq:loss} are accumulated over a two-stage rollout of the differentiable solver, illustrated in figure~\ref{fig:schematic}(b). Starting from the initial state $\boldsymbol{X}^0$, the flow is first advanced over a warm-up window of $N_1$ steps ($T_1$ flow-through times), with the closures active but gradients detached, which reduces the influence of the initial condition and lets the flow reach a statistically developed state. It is then rolled out over a gradient-tracking window of $N_2$ steps ($T_2$ flow-through times), over which the statistics are collected, spatially averaged in the homogeneous spanwise direction, and through which gradients are back-propagated. During training, $\boldsymbol{X}^0$ is taken from a fully developed turbulent field, which keeps the required warm-up short; otherwise $T_1$ would need to be long enough for the flow to become statistically developed, typically more than $20$ flow-through times. We use $N_1 = 400$ and $N_2 = 100$, corresponding to $T_1 = 1$ and $T_2 = 0.25$ flow-through times. This $T_1$ is sufficient to remove the initial-condition dependence, while $T_2$ yields converged statistics for the loss and also affects training stability under the differentiable rollout; an ablation on $T_1$ and $T_2$ is presented in \S \ref{sec:ablation_study}. Training is carried out for two boundary layers at $Re_\theta \in \{600,1200\}$, with the numerical settings summarized in table~\ref{tab:training_settings}.
\begin{table}
\begin{center}
\begin{tabular}{lccccccc}
\noalign{\smallskip}
$Re_{\theta}$ & Mesh & Domain / ($\theta$) & $\Delta t$ & $T_1$ & $N_1$ & $T_2$ & $N_2$ \\
\noalign{\smallskip}
\noalign{\smallskip}
600  & $64 \times 80 \times 128$ & $100 \times 80 \times 80$ & 0.25 & 1.0 & 400 & 0.25 & 100 \\
1200 & $64 \times 80 \times 128$ & $100 \times 80 \times 80$ & 0.25 & 1.0 & 400 & 0.25 & 100 \\
\noalign{\smallskip}
\end{tabular}
\caption{Training settings for the differentiable hybrid neural solver. All quantities are normalized by outer variables. Here, $T_1$ and $T_2$ are normalized by the flow-through time, while $N_1$ and $N_2$ are the corresponding numbers of time steps.}
\label{tab:training_settings}
\end{center}
\end{table}

The filtered Navier--Stokes equations~\eqref{eq:filtered_NS} are discretized with a finite-volume method and a fractional-step projection scheme. The convective term uses second-order linear interpolation with Rhie--Chow interpolation to suppress checkerboard pressure oscillations, and the viscous term a second-order central scheme. Time advancement is semi-implicit: the advection term is integrated with an explicit fourth-order Runge--Kutta method and the diffusion term with the Crank--Nicolson method. At the inlet, a recycling--rescaling method is used~\citep{lund1998generation}; a convective outflow condition is imposed at the outlet; the free-stream velocity is prescribed at the top boundary; and periodic conditions are applied in the spanwise direction. Further details and validation of the differentiable solver are given in \citet{fan2026diff}.

\vspace{0.3em}
\noindent\textbf{Training stability.}
Differentiating through long chaotic rollouts is known to be delicate, and numerical stability and rollout length are often critical for successful training~\citep{fan2025neural,franz2025pict,list2025differentiability}. To keep training stable, existing differentiable-physics studies typically resort to short gradient-tracking windows, offline pretraining of the network, curriculum strategies that gradually increase the rollout length, or selectively disabling gradients through certain operators~\citep{zhang2026differentiable,list2025differentiability,franz2025pict,fan2025neural,fan2024differentiable}. In the present framework, by contrast, stable training from scratch over a $100$-step gradient-tracking window is achieved without any pretraining. We attribute this to two factors: the semi-implicit time integration, which damps the fastest numerical modes, and the structure-constrained closures, whose network outputs are confined to a scaled Vreman coefficient and a wall-stress correction rather than the full stress fields. The framework is implemented in \texttt{JAX}, and gradient checkpointing (\texttt{jax.checkpoint}) is used so that gradients can be propagated through the long space--time rollout with manageable memory, in contrast to existing differentiable wall-model~\citep{zhang2026differentiable} and SGS-model~\citep{Kim2023GeneralizableDT} formulations that learn a single closure over shorter windows. The parameters are updated with the Adam optimizer. The full procedure is summarized in algorithm~\ref{alg:training}, and the network architectures and training hyper-parameters are given in appendix~\ref{sec:neural_net}.

\vspace{0.3em}
\noindent\textbf{Inference and deployment.}
Once trained, the closures are frozen and the hybrid model is deployed as a standard WMLES solver, as illustrated in figure~\ref{fig:schematic}(c). For a prescribed momentum-thickness Reynolds number $Re_\theta$, the solver advances an initial state $\boldsymbol{X}^0$ over a time horizon $T$ to the final state $\boldsymbol{X}^T$, exactly as a conventional solver would and with no further training. At inference the initial state is constructed from a mean profile based on the Blasius solution with superposed random perturbations, from which the flow transitions to turbulence. Although trained only at $Re_\theta \in \{600,1200\}$, the model is applied across a range of Reynolds numbers up to $Re_\theta = 6500$, as examined in \S\,\ref{sec:results}.

\begin{algorithm}
\small
\caption{Training of the differentiable hybrid neural framework}
\label{alg:training}
\begin{algorithmic}[1]
\STATE Initialize the trainable parameters $\boldsymbol{\theta} = [\theta_{\mathrm{sgs}}, \theta_{\mathrm{wall}}]$
\STATE Specify the training Reynolds numbers $\mathcal{R} = \{600, 1200\}$
\FOR{each training epoch}

    \STATE Distribute the training cases $Re_{\theta} \in \mathcal{R}$ across devices using \texttt{jax.pmap}
    \FOR{each $Re_{\theta} \in \mathcal{R}$ \textbf{in parallel}}
        \STATE Initialize $\boldsymbol{X}^0$ from a fully developed turbulent field
        \STATE Advance the solver over the warm-up window $T_1$ (gradients detached) to obtain $\boldsymbol{X}^{T_1}$
        \STATE Continue the differentiable rollout from $\boldsymbol{X}^{T_1}$ over the gradient-tracking window $T_2$
        \STATE Accumulate the turbulence statistics over $T_2$: $\langle \overline{u}_1^+ \rangle$, $\overline{u}_{i,\mathrm{rms}}^+$, $\overline{p}_{\mathrm{rms}}^+$ and $u_\tau$
        \STATE Compute the case-wise loss $\mathcal{L}_{Re_{\theta}}(\boldsymbol{\theta})$ and loss weights $w_k = 1/|\mathcal{L}_k|$ (detached) from equation~\eqref{eq:loss}

    \ENDFOR
    \STATE Aggregate the total loss $\mathcal{L}(\boldsymbol{\theta}) = \displaystyle\sum_{Re_{\theta} \in \mathcal{R}} \mathcal{L}_{Re_{\theta}}(\boldsymbol{\theta})$
    \STATE Back-propagate through the differentiable solver to obtain $\boldsymbol{g} = \nabla_{\boldsymbol{\theta}}\mathcal{L}(\boldsymbol{\theta})$
    \STATE Update $\boldsymbol{\theta}$ using the Adam optimizer
\ENDFOR
\end{algorithmic}
\end{algorithm}

\vspace{-2em}
\section{Numerical results}
\label{sec:results}

In this section, the performance of the proposed framework is assessed through a series of \textit{a posteriori} tests, summarized in table~\ref{tab:testing_settings}. Unless otherwise stated, each case is simulated for more than $40$ flow-through times, and the final $20$ are used to compute the turbulence statistics. For cases 1--8, the statistics are reported at the streamwise location where $Re_\theta$ attains the value listed in the table. Case 9 is treated differently: it is run in a very long domain with inlet Reynolds number $Re_{\theta,\mathrm{in}}=5000$, so that $Re_\theta$ grows in the streamwise direction, and statistics are reported at several downstream stations. Together, these cases probe the generalizability of the trained model across Reynolds numbers, computational domains, mesh resolutions and time steps.

The proposed differentiable hybrid neural model is denoted \textit{Hybrid-Joint}. We compare it against two conventional WMLES references, both using the equilibrium wall-stress model but different SGS closures. The first, denoted \textit{Baseline-Smagorinsky}, uses the standard Smagorinsky model with van Driest near-wall damping~\citep{van1956turbulent,nabae2025large}, a combination widely used in WMLES. The second, denoted \textit{Baseline-Vreman}, uses the standard constant-coefficient Vreman model~\citep{vreman2004eddy}. Further details of both baselines are given in appendix~\ref{sec:baseline}. Reference data are obtained from DNS, WRLES or wall-resolved-quality fine-mesh WMLES, depending on the case when DNS becomes prohibitively expensive; when turbulence statistics are compared, the reference fields are filtered and downsampled onto the same coarse mesh as \textit{Hybrid-Joint} solutions, and these processed data are referred to as the ground truth (GT). 

\begin{table}
    \centering
    \begin{tabular}{lcccccc}
        \hline
        Case & $Re_{\theta}$ & Mesh ($N_1\times N_2\times N_3$) & Domain / $(\theta)$ & $\Delta x_2^{+}$ & $\Delta t$ & $T$ \\
        \noalign{\smallskip}
        1 & 600  & $64 \times 80 \times 128$  & $100 \times 80 \times 80$  & 30    & 0.25  & 40 \\
        2 & 800  & $64 \times 80 \times 128$  & $100 \times 80 \times 80$  & 38    & 0.25  & 40 \\
        3 & 1200 & $64 \times 80 \times 128$  & $100 \times 80 \times 80$  & 55    & 0.10  & 40 \\
        4 & 1200 & $256 \times 80 \times 128$ & $400 \times 80^{\dagger} \times 80$ & 24--78 & 0.10  & 40 \\
        5 & 2400 & $64 \times 80 \times 128$  & $100 \times 80 \times 80$  & 102   & 0.05  & 40 \\
        6 & 3000 & $64 \times 80 \times 128$  & $100 \times 80 \times 80$  & 123   & 0.04  & 40 \\
        7 & 4000 & $64 \times 80 \times 128$  & $100 \times 80 \times 80$  & 160   & 0.03  & 40 \\
        8 & 5000 & $64 \times 80 \times 128$  & $100 \times 80 \times 80$  & 194   & 0.025 & 40 \\
        9 & $5000_{\mathrm{in}}$ & $512 \times 80 \times 128$ & $800 \times 80 \times 80$ & 194 & 0.025 & 10 \\
        \hline
    \end{tabular}
    \caption{Testing settings for the \textit{a posteriori} cases; the same settings are used for \textit{Hybrid-Joint} and both baselines. All quantities are normalized by outer variables, and $\Delta x_2^{+}$ denotes the wall-normal spacing of the first off-wall cell in wall units. Case~9 is a spatially developing boundary layer with inlet Reynolds number $Re_{\theta,\mathrm{in}}=5000$; its statistics are collected at several downstream stations. $^{\dagger}$Wall-normal grid stretching with a factor of $1.2$ is applied in the extended-domain case at $Re_{\theta}=1200$ (Case 4).}
    \label{tab:testing_settings}
\end{table}

The results are organized as follows. First, the hybrid model is tested at $Re_{\theta} = 600$ with a detailed comparison against DNS reference data and the two baseline models. Secondly, the model is tested over a broad Reynolds-number range, $Re_{\theta} \in [800, 5000]$, on the standard uniform coarse mesh, to examine its Reynolds-number generalizability, with $Re_{\theta} = 800$ and $Re_{\theta} = 5000$ examined in detail as representative interpolation and extrapolation testing cases at unseen Reynolds numbers, respectively. Finally, additional tests in enlarged computational domains and with modified mesh distributions, including the case at $Re_{\theta} = 1200$ in a long domain and the case at $Re_{\theta} = 5000$ in a very long domain, evaluate the sensitivity of the model to domain size, mesh distribution and resolution.

\subsection{\textit{A posteriori} test at the training Reynolds number}
\label{sec:Re600_results}

We first examine Case~1 at $Re_\theta = 600$, one of the two Reynolds numbers used in training. Although the Reynolds number is seen during training, this test is far from an in-sample check: the model is trained over a rollout of only $1.25$ flow-through times from a fully developed turbulent field, whereas here it is deployed from a randomly-perturbed Blasius profile and advanced autonomously for $40$ flow-through times. The test therefore probes whether closures optimized over a short window remain stable and statistically faithful over a rollout more than thirty times longer, and whether they can carry the flow through transition from an initial condition never seen in training.

\begin{figure}[htp!]
    \centerline{\includegraphics[width=0.9\textwidth]{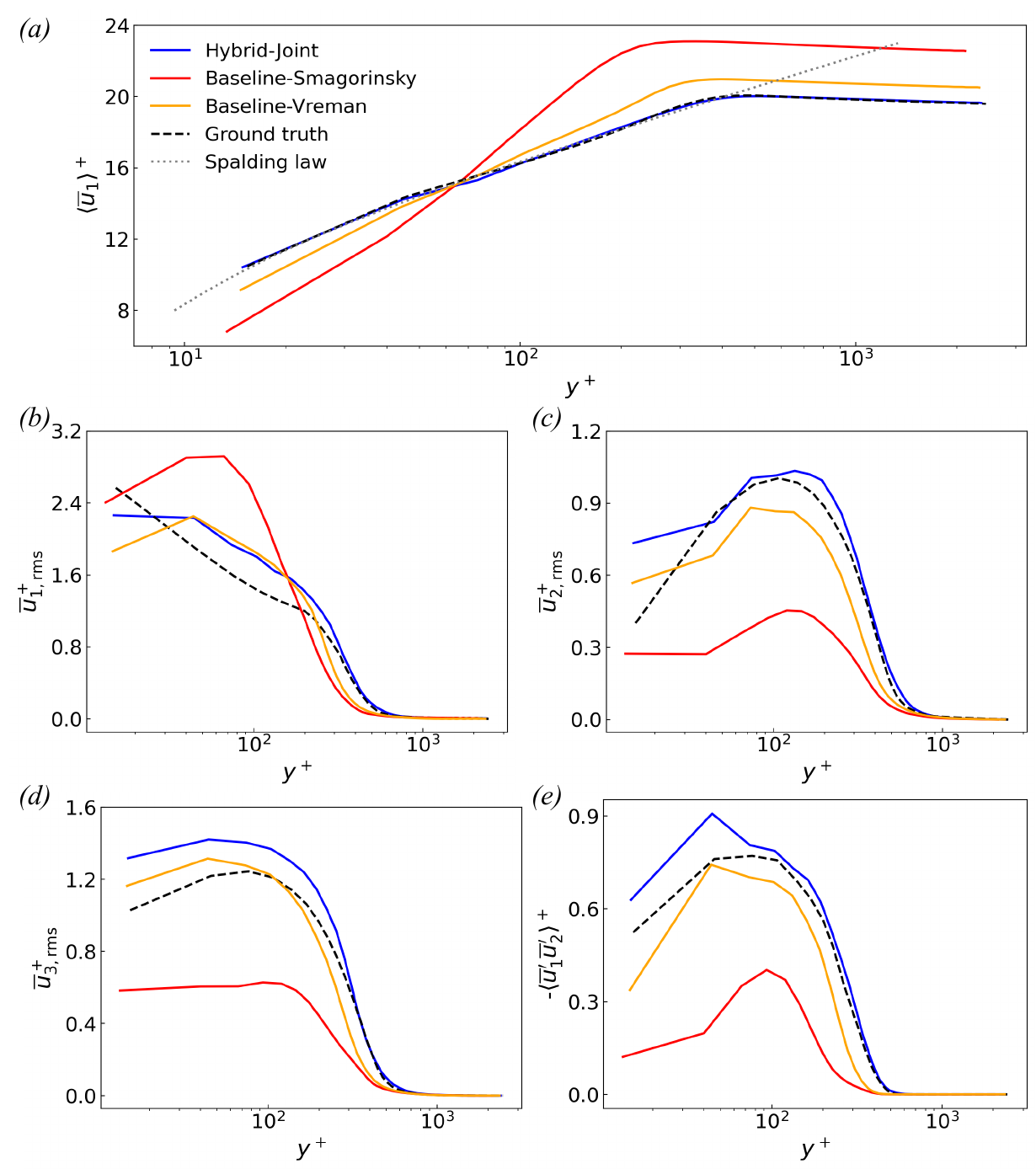}}
    \caption{Turbulence statistics for Case 1 ($Re_{\theta}=600$), compared with the two baselines and the ground truth (DNS). (a) Mean streamwise velocity profile. (b)--(d) Root-mean-square (RMS) velocity fluctuations $\overline{u}_{i,\mathrm{rms}}^{+}$. (e) Reynolds shear stress $-\langle\overline{u}_1'\overline{u}_2'\rangle^{+}$. All statistics are normalized by the friction velocity $u_{\tau}$.}
    \label{fig:Re600_stat}
\end{figure}

Figure~\ref{fig:Re600_stat} shows the first- and second-order turbulence statistics for Case~1 at $Re_{\theta}=600$. For the mean velocity profile in figure~\ref{fig:Re600_stat}(a), \textit{Hybrid-Joint} agrees closely with the reference data across the inner and outer layers and recovers the logarithmic region. \textit{Baseline-Smagorinsky} fails to recover the correct profile, with a pronounced mismatch in the logarithmic region, while \textit{Baseline-Vreman} improves substantially on it but still departs from the reference, slightly underpredicting the velocity in the buffer layer.

The mismatch of the baselines should not be interpreted as an inherent weakness of the baselines. Rather, it reflects the difficulty of combining a constant-coefficient SGS eddy-viscosity closure with an equilibrium wall-stress model on the very coarse mesh. WMLES of TBLs typically requires at least 20--30 grid points across the boundary-layer thickness~\citep{abkar2015influence}, whereas only about $10$ wall-normal points lie within $\delta_{99}$ here. The outer-layer motions are therefore strongly under-resolved, which corrupts the wall-stress prediction even with the third off-wall point used as the matching location, and the van Driest damping introduces further near-wall sensitivity. These effects lead to an underprediction of the friction velocity $u_{\tau}$ by approximately $12\%$, together with an incorrect outer-layer response, resulting in the observed log-layer mismatch~\citep{kawai2012wall,yang2017log}. The baseline profile improves at higher Reynolds numbers (\S\,\ref{sec:broadRe}), confirming that the implementation is consistent while its robustness on this coarse grid is limited. That \textit{Hybrid-Joint} recovers the correct profile under the same numerical setting demonstrates its ability to compensate for the coupled effects of SGS modelling, wall modelling and numerical under-resolution.

The r.m.s.\ velocity fluctuations and the Reynolds shear stress are shown in figures~\ref{fig:Re600_stat}(b)--(e). \textit{Hybrid-Joint} agrees well with the ground truth overall, with a slight overprediction of the near-wall peaks of $\overline{u}_{1,\mathrm{rms}}^{+}$, $\overline{u}_{3,\mathrm{rms}}^{+}$ and $-\langle\overline{u}_1'\overline{u}_2'\rangle^{+}$. \textit{Baseline-Smagorinsky} is markedly over-dissipative: it overpredicts $\overline{u}_{1,\mathrm{rms}}^{+}$ while strongly underpredicting the wall-normal and spanwise fluctuations and the Reynolds shear stress. \textit{Baseline-Vreman} follows the reference trends for all components but with lower accuracy than \textit{Hybrid-Joint}, consistently underpredicting the fluctuation peaks. Such discrepancies are common in WMLES and are known to depend on the SGS closure, grid resolution and numerical dissipation~\citep{fan2025near,fowler2022lagrangian}.

\begin{figure}[htp!]
    \centerline{\includegraphics[width=\textwidth]{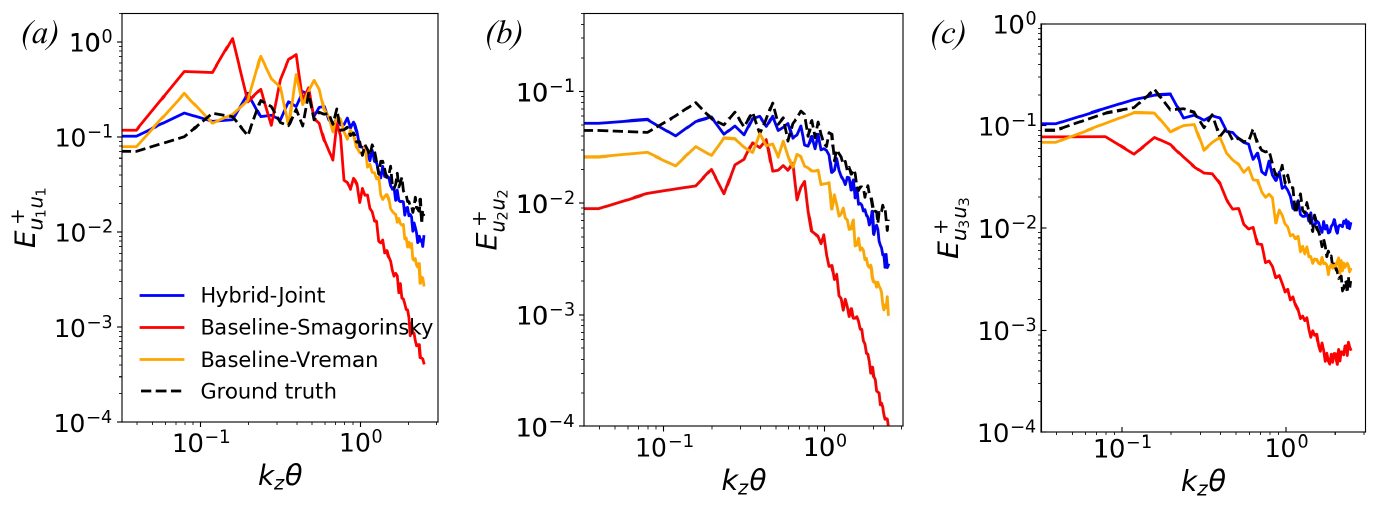}}
    \caption{Spanwise wavenumber energy spectra of the velocity components for Case 1 ($Re_{\theta}=600$) at $y^{+}=138$. (a) Streamwise velocity. (b) Wall-normal velocity. (c) Spanwise velocity. The spectra are normalized by $u_{\tau}$. For consistency, the GT is filtered at the cut-off wavenumber associated with the same mesh.}
    \label{fig:600_energy}
\end{figure}

\begin{figure}[htp!]
    \centerline{\includegraphics[width=0.9\textwidth]{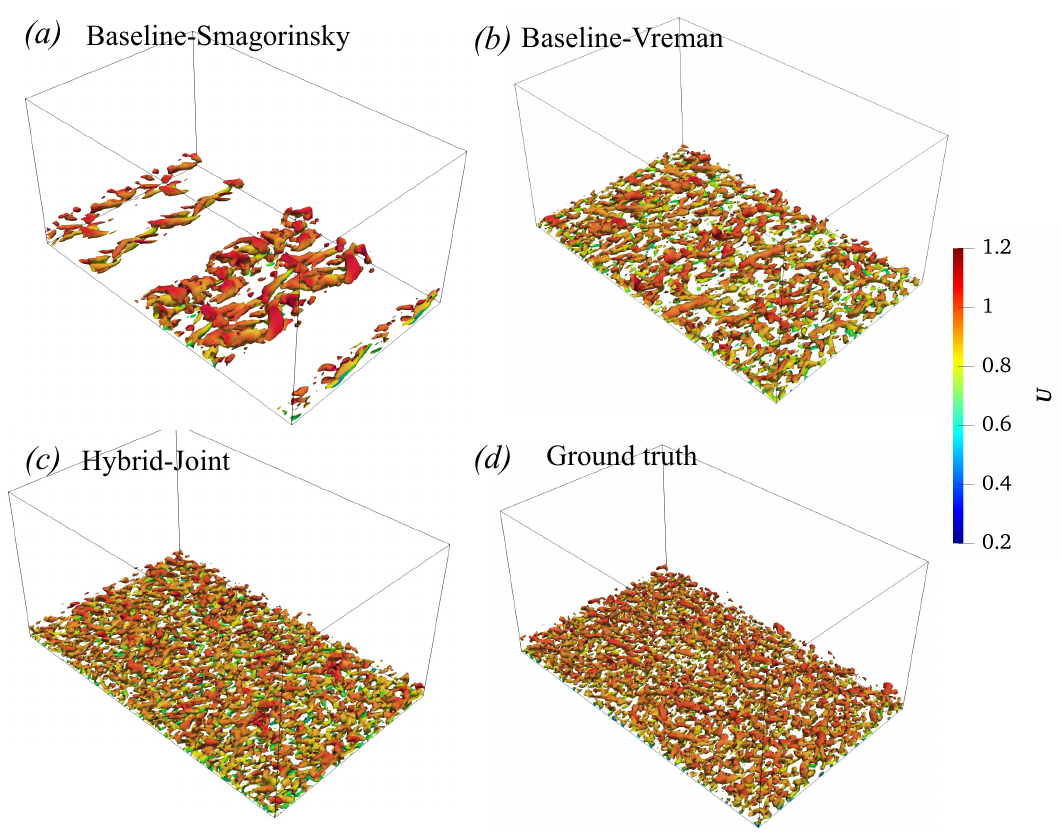}}
    \caption{Comparison of the instantaneous velocity field predicted by different models for Case 1 ($Re_{\theta}=600$). (a) \textit{Baseline-Smagorinsky} (b) \textit{Baseline-Vreman} (c)  \textit{Hybrid-Joint} (d) Ground truth, obtained by filtering the DNS solution and downsampling it onto the same coarse mesh. The isosurfaces are defined by the $Q$ criterion at $Q(\delta_{99}/U_{\infty})^2=10^{-3}$, and the colour denotes the velocity magnitude.}
    \label{fig:600_contour}
\end{figure}

Figure~\ref{fig:600_energy} shows the one-dimensional spanwise wavenumber spectra of the three velocity components at $y^{+}=138$. \textit{Hybrid-Joint} accurately reproduces the reference spectra across the resolved wavenumber range, indicating that the distribution of resolved kinetic energy across scales is well preserved; the spectral shape is recovered even though the spectra are not part of the training loss, being constrained only indirectly through the r.m.s.\ fluctuations. \textit{Baseline-Smagorinsky} underpredicts the energy over a broad range of wavenumbers, most severely for the wall-normal and spanwise components, while \textit{Baseline-Vreman} predictions lie closer to the reference values but remain less accurate than \textit{Hybrid-Joint}. The instantaneous fields visualized by the $Q$-criterion, shown in figure~\ref{fig:600_contour}, are consistent with these statistics: \textit{Hybrid-Joint} recovers near-wall turbulent structures in close agreement with the ground truth, \textit{Baseline-Vreman} captures fewer structures, and \textit{Baseline-Smagorinsky} retains only the largest near-wall structures, filtering out most of the smaller-scale activity.

\subsection{\textit{A posteriori} tests at unseen Reynolds numbers}
\label{sec:broadRe}

Having been trained only at $Re_\theta = 600$ and $1200$, the model is now applied across $Re_\theta \in [800, 5000]$, spanning an interpolation case ($Re_\theta = 800$) and extrapolation to more than four times the highest training Reynolds number ($Re_\theta = 5000$). All cases use the same coarse uniform mesh, so that the near-wall resolution in inner scale coarsens from $\Delta x_2^+ \approx 38$ to $194$ as the Reynolds number increases (table~\ref{tab:testing_settings}).

\begin{figure}[htp!]
    \centerline{\includegraphics[width=\textwidth]{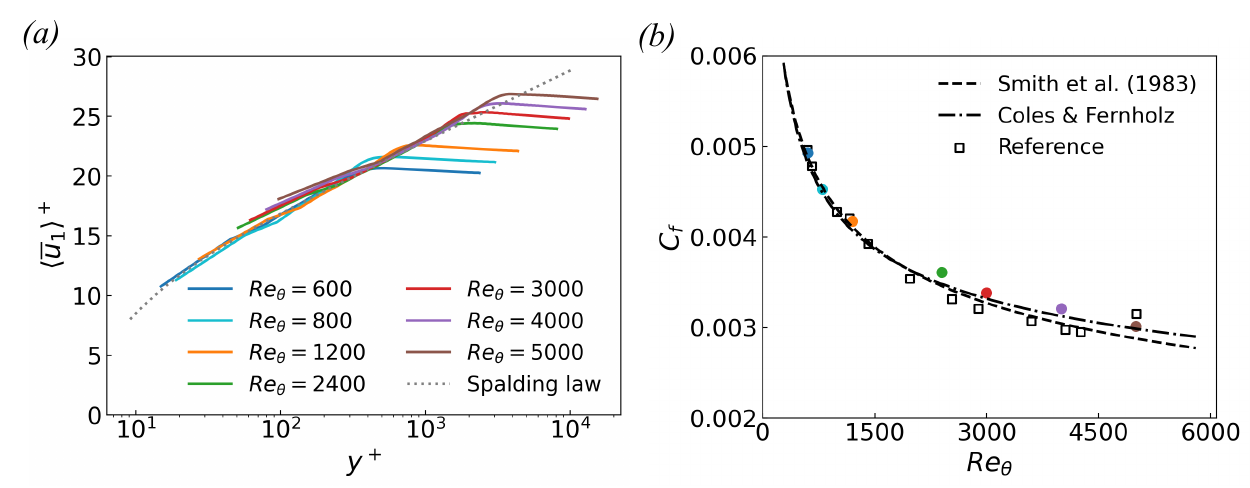}}
    \caption{Assessment of the model performance over the Reynolds-number range $Re_{\theta}=600$--5000. (a) Mean streamwise velocity profiles, compared with the law of the wall. (b) Skin-friction coefficient as a function of the momentum-thickness Reynolds number, both evaluated from the simulated flow, compared with the empirical correlations of \citet{smits1983low} and \citet{nagib2023reynolds}, and with the reference data of \citet{fernholz1996incompressible}.}
    \label{fig:all_mean}
\end{figure}

Figure~\ref{fig:all_mean}(a) shows the mean velocity profiles over the range. A clear logarithmic region develops at every Reynolds number, in agreement with the law of the wall, with a slight departure only at the first off-wall point. This agreement is a genuine test of the model: unlike conventional equilibrium wall models, which invert an assumed law of the wall at a matching location, the present wall closure never matches the resolved velocity to a law of the wall. The logarithmic behaviour is therefore not imposed but emerges from the simulated flow.
Figure~\ref{fig:all_mean}(b) shows the corresponding skin-friction coefficient against the momentum-thickness Reynolds number, both evaluated \textit{a posteriori} from the simulated flow. The predictions follow the expected $Re_\theta$ scaling across the whole range and lie within the scatter of the reference data of \citet{fernholz1996incompressible} and the empirical correlations of \citet{smits1983low} and \citet{nagib2023reynolds}. Because $Re_\theta$ follows from the momentum thickness, and hence from the mean velocity profile shaped jointly by the SGS and wall closures, reproducing this relation as the Reynolds number increases requires the two closures to remain consistent with one another.

Two representative cases are then examined in detail against the ground truth: the interpolation case at $Re_\theta = 800$ (Case~2) and the extrapolation case at $Re_\theta = 5000$ (Case~8). The first- and second-order turbulence statistics are reported in figures~\ref{fig:Re800_stat} and \ref{fig:Re5000_stat} in Appendix~\ref{sec:additional_results}. In both cases the behaviour is similar to that at $Re_\theta = 600$: \textit{Hybrid-Joint} agrees closely with the ground truth and clearly outperforms \textit{Baseline-Smagorinsky}, while \textit{Baseline-Vreman} follows the reference reasonably well and \textit{Hybrid-Joint} remains somewhat closer, most visibly in $\overline{u}_{3,\mathrm{rms}}^{+}$ and in the outer part of the mean profile. That the agreement is retained at $Re_\theta = 5000$ indicates that extrapolating more than fourfold beyond the training range does not degrade the prediction.

\begin{figure}[htp!]
    \centerline{\includegraphics[width=\textwidth]{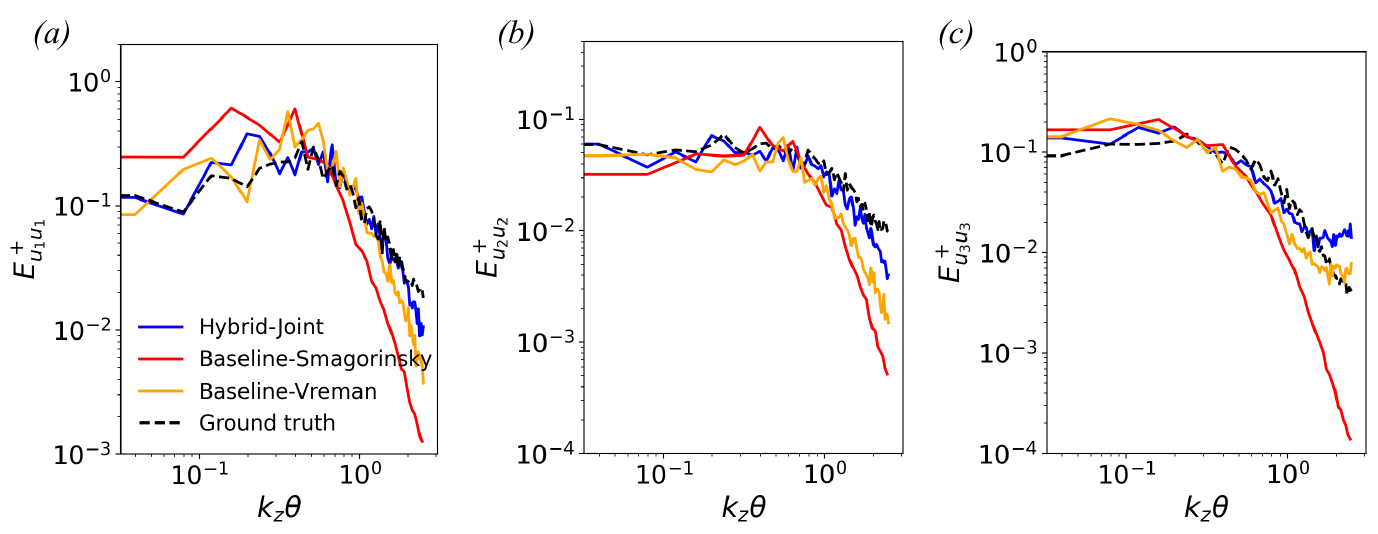}}
    \caption{Spanwise wavenumber energy spectra of the velocity components for Case 2 ($Re_{\theta}=800$) at $y^{+}=170$. (a) Streamwise velocity. (b) Wall-normal velocity. (c) Spanwise velocity. The spectra are normalized by $u_{\tau}$. For consistency, the GT is filtered at the cut-off wavenumber associated with the same mesh.}
    \label{fig:Re800_energy}
\end{figure}
\begin{figure}[htp!]
    \centerline{\includegraphics[width=\textwidth]{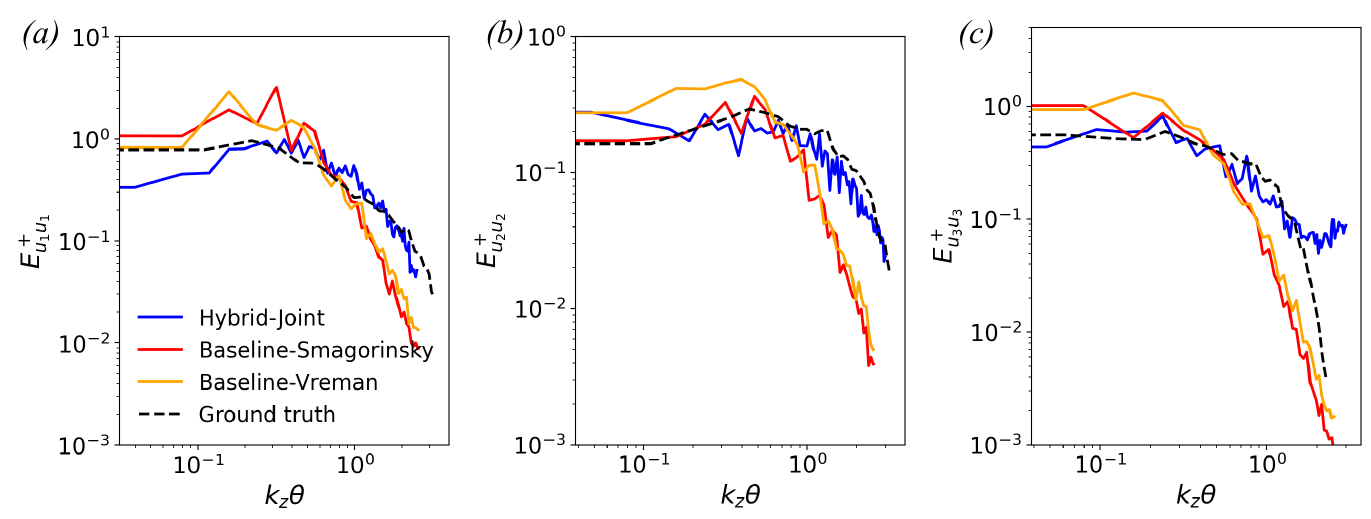}}
    \caption{Spanwise wavenumber energy spectra of the velocity components for Case 8 ($Re_{\theta}=5000$) at $y^{+}=600$. (a) Streamwise velocity. (b) Wall-normal velocity. (c) Spanwise velocity. The spectra are normalized by $u_{\tau}$. For consistency, the GT is filtered at the cut-off wavenumber associated with the same mesh.}
    \label{fig:Re5000_energy}
\end{figure}
The corresponding spanwise wavenumber spectra are shown in figures~\ref{fig:Re800_energy} and \ref{fig:Re5000_energy}. At $Re_\theta = 800$, \textit{Hybrid-Joint} follows closely the reference spectra across the resolved wavenumber range for all three components, and \textit{Baseline-Vreman} remains reasonably close, while \textit{Baseline-Smagorinsky} overpredicts the streamwise energy $E^+_{u_1u_1}$ at low wavenumbers and is excessively dissipative at high wavenumbers. At $Re_\theta = 5000$, \textit{Hybrid-Joint} continues to track the reference, whereas \textit{Baseline-Vreman} deteriorates markedly: it overpredicts the energy at low and intermediate wavenumbers for all three components and, like \textit{Baseline-Smagorinsky}, falls off too rapidly at high wavenumbers. The distribution of resolved energy across scales is therefore preserved by \textit{Hybrid-Joint} even four times beyond the training range, where both fixed-coefficient closures degrade. A slight excess in the high-wavenumber tail of the spanwise component $E^+_{u_3u_3}$ is observed for \textit{Hybrid-Joint} and appears consistently across the tested cases.

\subsection{Transferability to a different mesh distribution}
\label{sec:mesh}

The model is trained in the loop on a uniform grid in a domain of $100\theta$. We now examine whether the trained closures transfer when both the domain and the wall-normal grid distribution are changed. The test is performed at $Re_\theta = 1200$ with two configurations: Case~3, which uses the same uniform grid and domain as in training, and Case~4, which uses a domain four times longer ($400\theta$) together with wall-normal stretching at a factor of $1.2$, keeping the streamwise spacing unchanged (table~\ref{tab:testing_settings}). The stretching redistributes the wall-normal points, refining the first cell from $\Delta x_2^+ \approx 55$ to $\approx 24$ while coarsening the outer layer to $\approx 78$. Case~4 therefore departs from the training configuration in both domain length and mesh distribution simultaneously. The reference DNS data are obtained in a domain comparable to that of Case~3, so the comparison for Case~4 is made at matched $Re_\theta$ but in a longer domain than the reference.

\begin{figure}[htp!]
    \centerline{\includegraphics[width=\textwidth]{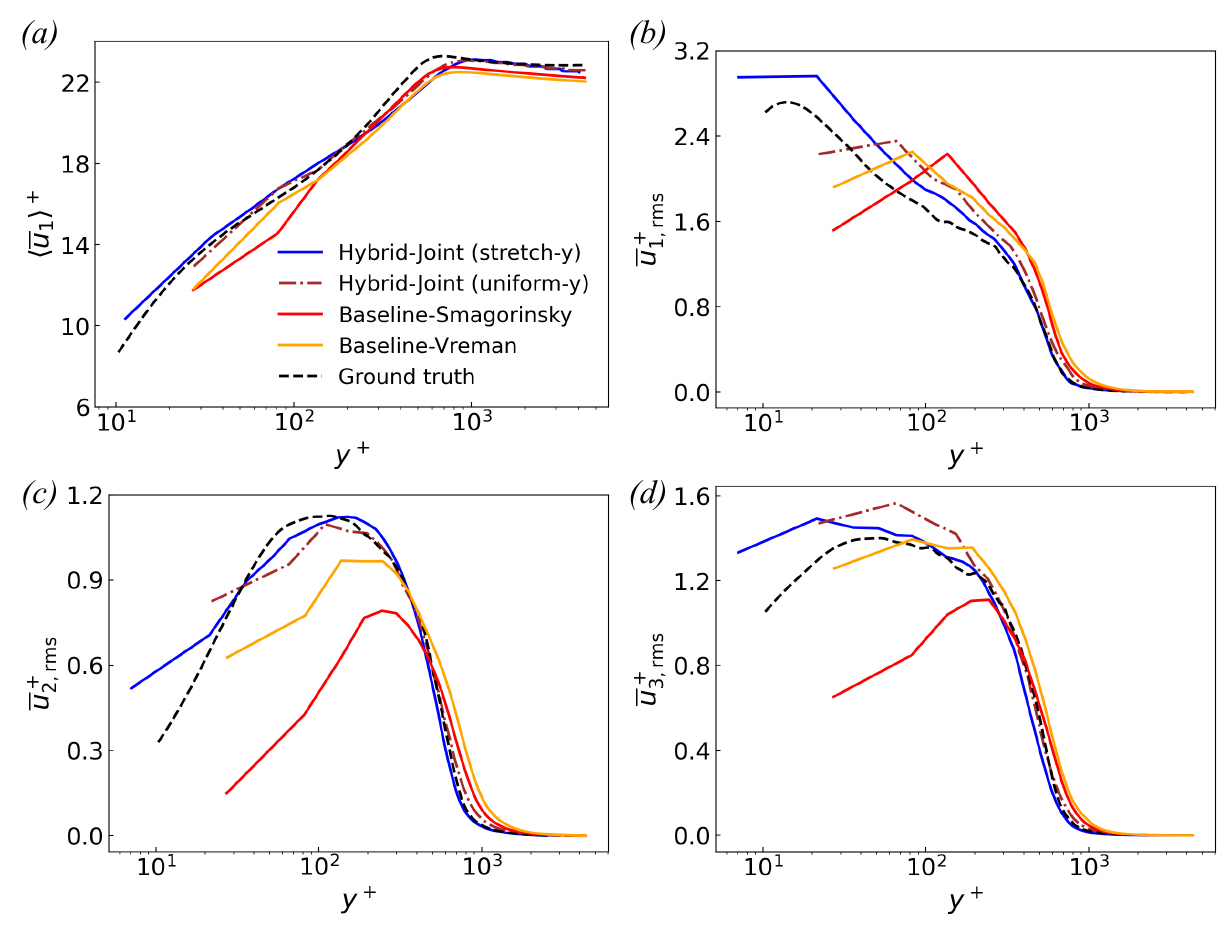}}
    \caption{Turbulence statistics at $Re_{\theta}=1200$ for Case~3 (uniform grid, $100\theta$ domain) and Case~4 (wall-normal stretched grid, $400\theta$ domain), compared with the two baselines and the ground truth (DNS). (a) Mean streamwise velocity profile. (b)--(d) Root-mean-square velocity fluctuations $\overline{u}_{i,\mathrm{rms}}^{+}$. All statistics are normalized by the friction velocity $u_{\tau}$. The reference and baselines are obtained in the same domain of Case~3.}
    \label{fig:Re1200_stat}
\end{figure}

Figure~\ref{fig:Re1200_stat}(a) compares the mean velocity profiles. \textit{Hybrid-Joint} follows the ground truth on both configurations and captures the logarithmic region, whereas both baselines underpredict the velocity through the buffer layer. On the stretched grid the profile extends to $y^+ \approx 10$, inside a region that the training mesh never resolved, and the prediction there remains close to the reference.
The r.m.s.\ velocity fluctuations are shown in figures~\ref{fig:Re1200_stat}(b)--(d). The wall-normal and spanwise components are reproduced well in both configurations and are substantially closer to the reference than the baselines, which are markedly over-dissipative throughout, especially for Smagorinsky baseline. The streamwise fluctuations in panel~(b) show the largest difference between the two configurations: on the longer stretched-grid domain, the near-wall peak of $\overline{u}_{1,\mathrm{rms}}^+$ exceeds the reference, whereas on the matched short domain it follows it closely. Since $\overline{u}_{1,\mathrm{rms}}^+$ is the component most sensitive to the large-scale streamwise motions, and since the reference is computed in the shorter domain, this difference cannot be attributed to the mesh distribution alone: the longer domain accommodates streamwise scales that the reference domain truncates, which is expected to raise the streamwise fluctuation level.

The model thus remains stable and accurate on a domain and grid distribution absent from training, which reflects two features of the framework: the closures are convolutional and act on the resolved field rather than on a fixed grid, and they are trained $\textit{a posteriori}$, so that they respond to the fully discretized solver rather than to a prescribed target.

\subsection{A spatially developing boundary layer in a very long domain}
\label{sec:case9}

The final and most demanding test is Case~9, a spatially developing boundary layer in a domain $800\theta_{\mathrm{in}}$ long, eight times the training domain, with inlet Reynolds number $Re_{\theta,\mathrm{in}} = 5000$. The boundary layer grows continuously downstream, so that $Re_\theta$ increases from its inlet value to approximately $6500$ at the outlet, and the model must sustain this development on the coarsest mesh considered here, without the statistics being anchored to a single streamwise station.

\begin{figure}[htp]
    \centerline{\includegraphics[width=\textwidth]{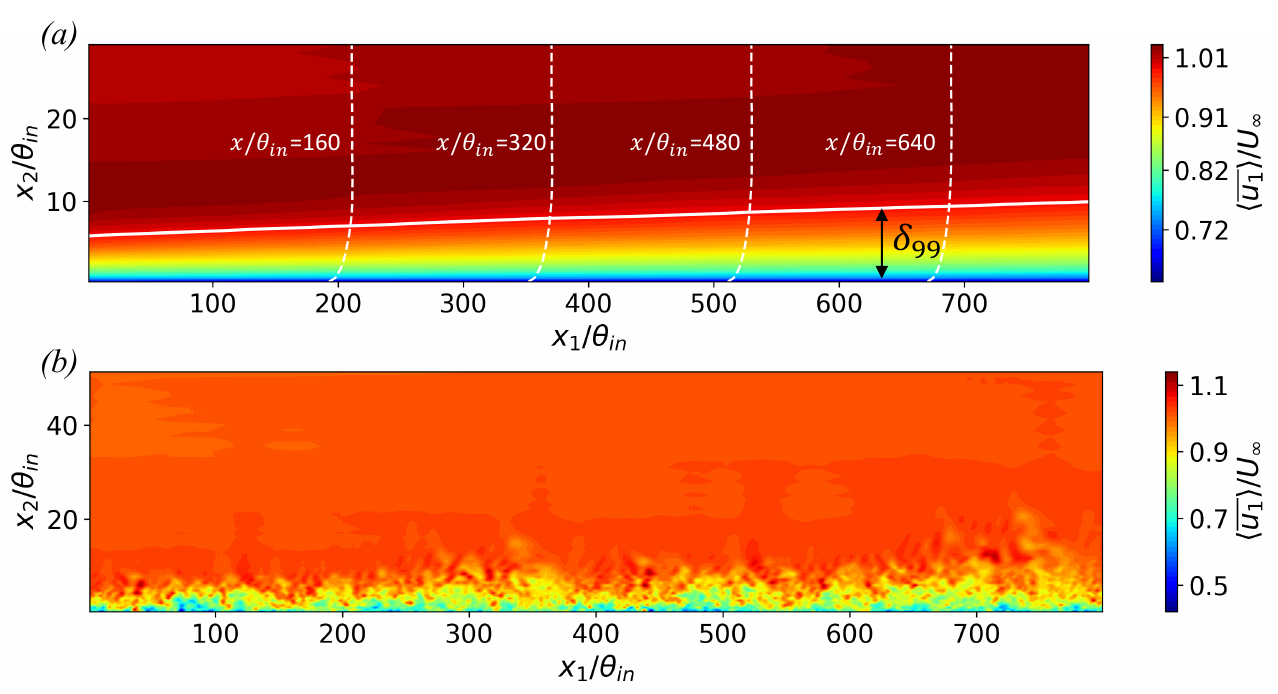}}
    \caption{Streamwise velocity for Case 9 ($Re_{\theta,\mathrm{in}}=5000$). (a) Mean streamwise velocity contour, boundary-layer thickness, and velocity profiles at selected streamwise locations. For clarity, the velocity profiles are amplified by a factor of 50. (b) Instantaneous streamwise velocity contour.}
    \label{fig:5000_long_contour}
\end{figure}

Figure~\ref{fig:5000_long_contour}(a) shows the mean streamwise velocity field together with the boundary-layer thickness. The downstream growth of $\delta_{99}$ is captured smoothly across the whole domain, and the mean profiles at the selected stations retain their shape as the layer thickens. The instantaneous field in figure~\ref{fig:5000_long_contour}(b) shows rich near-wall structures sustained over the full domain length despite the very coarse resolution, whereas WMLES baselines oversmooth them (figure~\ref{fig:Re5000_long_classic_contour}).

\begin{figure}[htp]
    \centerline{\includegraphics[width=\textwidth]{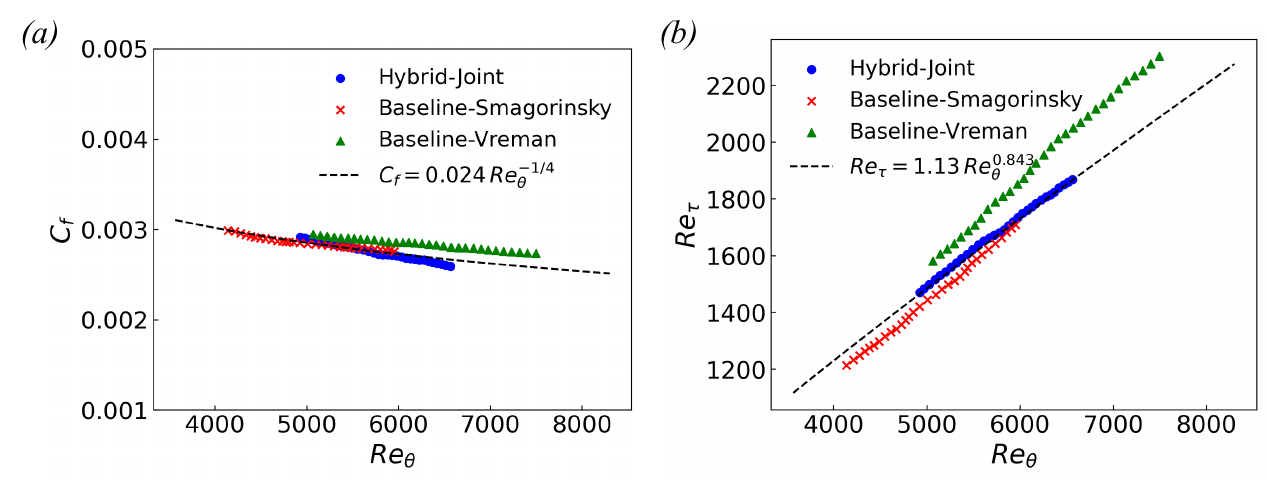}}
    \caption{Streamwise development of wall-friction quantities for Case 9 ($Re_{\theta,\mathrm{in}}=5000$). (a) Skin-friction coefficient along the streamwise direction. (b) Predicted friction Reynolds number $Re_{\tau}$ as a function of the Reynolds number based on momentum thickness, $Re_{\theta}$.}
    \label{fig:5000_long_cf}
\end{figure}

A quantitative assessment is given in figure~\ref{fig:5000_long_cf}. Panel~(a) shows the streamwise development of the skin-friction coefficient against the local $Re_\theta$, both evaluated from the simulated flow, over the same simulated distance and duration for all three models. All three follow the expected $C_f$--$Re_\theta$ trend, but reach markedly different states over that distance. \textit{Hybrid-Joint} starts at the prescribed inlet value $Re_\theta \approx 5000$ and develops to $\approx 6500$. \textit{Baseline-Smagorinsky} starts substantially below the target, near $Re_\theta \approx 4100$ ($18\%$ low), consistent with its momentum thickness being corrupted from the outset by excessive SGS dissipation (\S\,\ref{sec:learned}). \textit{Baseline-Vreman}, by contrast, starts at the correct inlet value but develops much faster than \textit{Hybrid-Joint}, reaching $Re_\theta \approx 8000$ over the same distance: its fixed, comparatively low dissipation lets the momentum thickness grow at an unphysically high rate.
In panel~(b), the friction Reynolds number $Re_\tau = u_\tau \delta_{99}/\nu$ involves the boundary-layer thickness, which no correlation prescribes, so the $Re_\tau$--$Re_\theta$ relation provides an independent check on the development of the layer. \textit{Hybrid-Joint} follows the correlation of \citet{schlatter2010assessment} closely over the whole range. Both baselines deviate, in opposite directions and to different degrees: \textit{Baseline-Smagorinsky} underpredicts $Re_\tau$ at a given $Re_\theta$, while \textit{Baseline-Vreman} overpredicts it more severely, consistent with its faster and unphysical growth of $Re_\theta$ in panel~(a). The two panels together show that only the jointly trained closures keep the inner scaling, reflected in $Re_\tau$, and the integral development of the layer, reflected in $Re_\theta$, simultaneously consistent with the reference correlations; each WMLES baseline instead reaches an incorrect boundary-layer state.

\section{Discussion}
\label{sec:discussion}

\subsection{What is learned by the \textit{Hybrid-Joint} model?}
\label{sec:learned}
In this section, we analyse the individual components of the proposed \textit{Hybrid-Joint} framework to clarify what is learned by the neural SGS and wall models when only turbulence statistics are used as training targets. This analysis provides insight into how the accuracy and generalizability observed in the \textit{a posteriori} tests arise from the design of the framework, and it identifies a structural consequence of the eddy-viscosity form that the neural SGS closure inherits.

\subsubsection{Neural SGS model} 
\label{sec:sgs_analysis}
\begin{figure}[htp!]
    \centerline{\includegraphics[width=\textwidth]{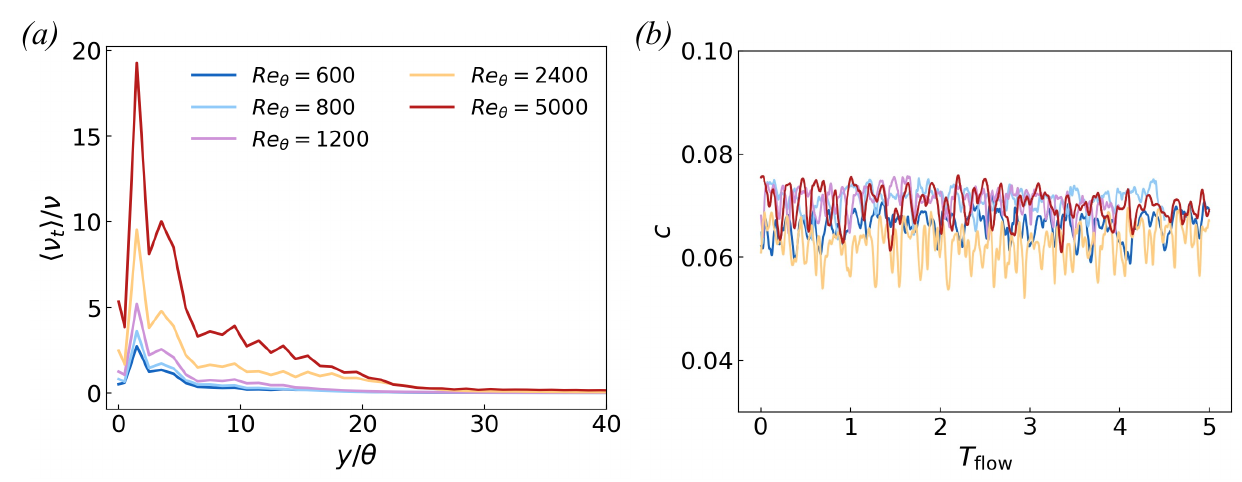}}
    \caption{Prediction of the neural SGS and wall models. (a) Mean subgrid viscosity $\nu_t$ over wall-normal distance, combining the interior SGS contribution and the wall contribution $\langle\nu_{t}^{\mathrm{wall}}\rangle$ (first point). (b) Time history of the wall-parallel-averaged Vreman coefficient $c$.}
    \label{fig:all_vis}
\end{figure}

The neural SGS and wall closures each augment the near-wall viscosity through independent mechanisms: the SGS closure sets the eddy viscosity $\nu_t$ at the interior cell centres (\S\,\ref{sec:sgs_closure}), while the wall closure sets $\nu_t^{\mathrm{wall}}$ through the modified boundary condition (\S\,\ref{sec:wall_closure}).  Although structurally separate, the two are optimized jointly against the same resolved field, so that the near-wall viscous augmentation they produce is coupled through training rather than through a shared functional form. Figure~\ref{fig:all_vis}(a) shows both contributions together as a function of wall-normal distance for different Reynolds numbers, with the first point corresponding to $\langle\nu_{t}^{\mathrm{wall}}\rangle$ and the remaining points to the interior values, illustrating how the two closures jointly augment the near-wall viscosity.

\begin{figure}[htp!]
    \centerline{\includegraphics[width=\textwidth]{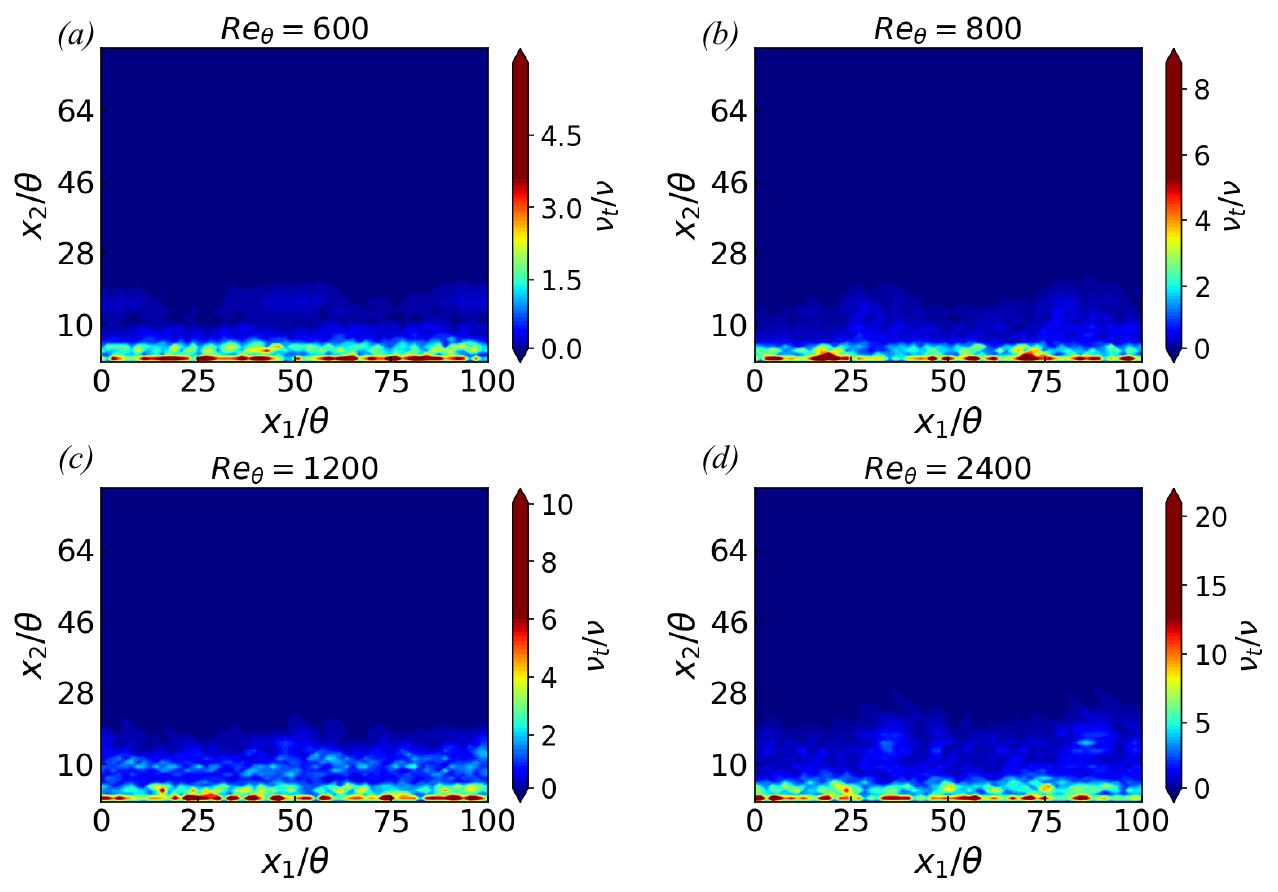}}
    \caption{Instantaneous subgrid scale eddy viscosity $\nu_t$ predicted by the neural SGS model. (a) Case 1, $Re_\theta=600$. (b) Case 2, $Re_\theta=800$. (c) Case 3, $Re_\theta=1200$. (d) Case 5, $Re_\theta=2400$.}
    \label{fig:instantaneous_vis}
\end{figure}

The predicted eddy viscosity is much larger than the molecular viscosity within the boundary layer, with a pronounced peak a few cells off the wall, indicating substantial SGS-mediated energy transfer from the resolved to the unresolved scales. The magnitude of $\langle\nu_t\rangle/\nu$ within the boundary layer increases with $Re_\theta$, consistent with the expected Reynolds-number dependence of turbulent mixing, and decays to zero in the free stream in every case. Similar near-wall behaviour has been reported for the conventional Vreman model in turbulent channel flow~\citep{whitmore2020requirements}, although in that case $\langle\nu_{t}^{\mathrm{wall}}\rangle=0$ owing to the Neumann condition imposed at the wall, and the peak occurs at the first off-wall cell rather than the second. The instantaneous fields in figure~\ref{fig:instantaneous_vis} confirm that the SGS dissipation remains confined within the boundary layer, a behaviour naturally induced by the Vreman form.

Figure~\ref{fig:all_vis}(b) shows the time history of the Vreman coefficient $c$, obtained by averaging the raw network output $c_m(\boldsymbol{x},t)$ over the entire spatial domain. At every Reynolds number, averaged $c$ fluctuates persistently in time about an approximately constant level, without any drift or slow trend over the rollout, and both the fluctuation amplitude and the mean level are broadly similar across $Re_\theta=600$--5000. The coefficient shows only a weak dependence on $Re_\theta$ and no clear monotonic trend, indicating that the Reynolds-number dependence of the eddy viscosity in figure~\ref{fig:all_vis}(a) arises mainly through the predicted velocity gradients entering the Vreman form, rather than through $c$ itself. The resulting level of this averaging is close to $c\approx0.07$, the value derived for homogeneous isotropic turbulence~\citep{zhou2025effect} and used as the fixed coefficient in \textit{Baseline-Vreman}. This is an emergent outcome of training on turbulence statistics, not a value the network was steered toward: nothing in the loss or the architecture references this constant, and the raw output $c_m$ is a free spatiotemporal field with no constraint on its sign or magnitude. The trained network contributes less through its mean output than through its spatio-temporal structure: unlike a constant-coefficient model, $c$ retains wall-normal variation, and it fluctuates with the resolved flow rather than staying fixed. The framework thus identifies a coefficient that is, on average, close to the classical constant, while retaining the flexibility to adapt to the local, instantaneous flow -- a distinction reflected in the improved \textit{a posteriori} performance of \textit{Hybrid-Joint} over \textit{Baseline-Vreman}, especially at $Re_\theta$ beyond the training range (\S\,\ref{sec:broadRe}).

We further perform an energy-transfer analysis to examine whether the model captures both forward scatter and backscatter. The filtered kinetic-energy equation is obtained by multiplying the filtered momentum equation~\eqref{eq:filtered_NS} by $\overline{u}_i$,
\begin{equation}
\frac{\partial E_f}{\partial t}
+ \overline{u}_j \frac{\partial E_f}{\partial x_j}
+ \frac{1}{\rho}\frac{\partial \overline{u}_i \overline{p}}{\partial x_i}
+ \frac{\partial \overline{u}_i \tau_{ij}}{\partial x_j}
- 2\nu \frac{\partial \overline{u}_i \overline{S}_{ij}}{\partial x_j}
= -\epsilon_f - \Pi ,
\end{equation}
where $E_f=\overline{u}_i\overline{u}_i/2$ is the kinetic energy of the filtered velocity field and $\epsilon_f=2\nu \overline{S}_{ij}\overline{S}_{ij}$ is the viscous dissipation. The SGS energy-transfer rate is
\begin{equation}
    \Pi = -\tau_{ij}\overline{S}_{ij}.
\end{equation}
A positive $\Pi$ corresponds to forward scatter, the transfer of kinetic energy from the resolved to the unresolved scales, while a negative $\Pi$ indicates backscatter. The mean SGS transfer is expected to be predominantly forward, but instantaneous backscatter events are known to occur locally and are relevant for assessing the learned SGS closure.

\begin{figure}[htp!]
    \centerline{\includegraphics[width=\textwidth]{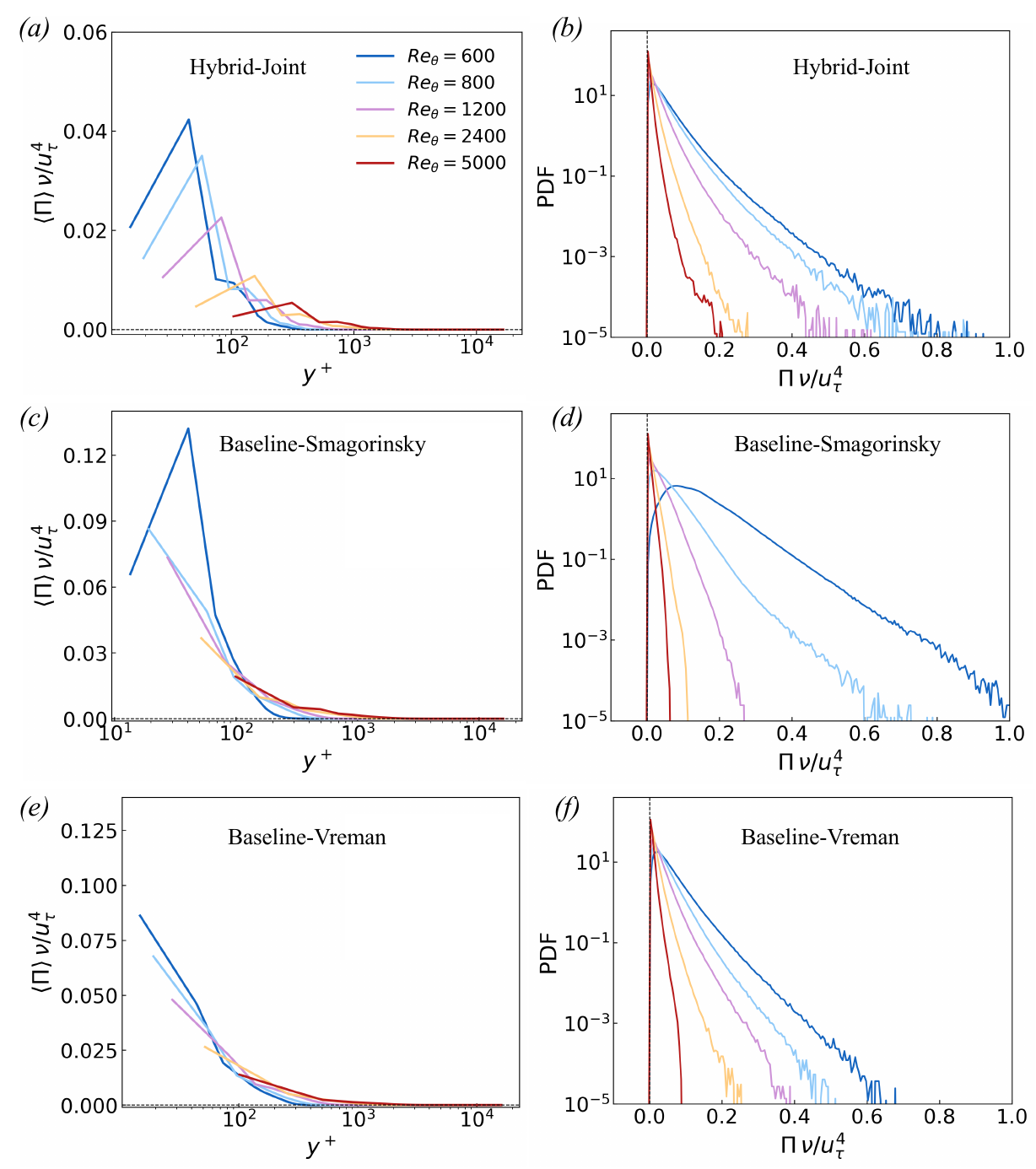}}
    \caption{Energy transfer between the resolved and subgrid scales. (a,c,e) Wall-normal profiles of the mean SGS energy-transfer rate $\langle \Pi \rangle$ for \textit{Hybrid-Joint}, \textit{Baseline-Smagorinsky} and \textit{Baseline-Vreman}, respectively. (b,d,f) Probability density functions of the instantaneous energy-transfer rate $\Pi$ at $x_2/\delta_{99}=0.2$.}
    \label{fig:energy_transfer}
\end{figure}

Figure~\ref{fig:energy_transfer}(a,c,e) compares the mean SGS energy-transfer rate $\langle\Pi\rangle$ across the three models. In every case $\langle\Pi\rangle$ is positive, confirming forward scatter on average. Both baselines produce substantially larger peaks than \textit{Hybrid-Joint}, with \textit{Baseline-Smagorinsky} the most dissipative of the three; \textit{Baseline-Vreman}, despite sharing the same functional form and a coefficient close on average to that learned by \textit{Hybrid-Joint}, still overpredicts the energy transfer by roughly a factor of two near the wall. This reinforces that the fixed coefficient, not the Vreman form itself, is responsible for the excess dissipation, and that the wall-normal structure and flow-dependent fluctuations retained by the learned coefficient are what bring $\langle\Pi\rangle$ down to a level, $\Pi\nu/u_\tau^4$, consistent with \textit{a priori} analyses of wall-bounded turbulence~\citep{piomelli1996subgrid}.

Figure~\ref{fig:energy_transfer}(b,d,f) shows the corresponding PDFs of the instantaneous $\Pi$ at $x_2/\delta_{99}=0.2$. All three models produce distributions confined to non-negative values: none represents backscatter in the present \textit{a posteriori} simulations. However, backscatter is known to occur in TBLs, more prominently in the buffer layer and, more weakly but still present, in the logarithmic region~\citep{piomelli1991subgrid,piomelli1996subgrid,chedevergne2025characterisation,wang2021energy}. Consistent with this, the \textit{a priori} analysis at $Re_\theta=600$ in figure~\ref{fig:Re600_priori_enenrgy} shows a clear negative tail of $\Pi$ at $x_2/\delta_{99}=0.2$ in the filtered reference field, confirming that local backscatter is present in the flow even though none of the three closures represents it.

This absence follows from the eddy-viscosity form shared by all three closures: with $\tau_{ij}=-2\nu_t\overline{S}_{ij}$, the transfer rate $\Pi = 2\nu_t\,\overline{S}_{ij}\overline{S}_{ij}$ is pointwise proportional to $\nu_t$, so its sign is fixed entirely by the sign of the eddy viscosity, independent of the resolved strain field. For the two baselines, the coefficient is fixed and positive by construction, so $\Pi\ge0$ is guaranteed. The neural SGS closure in \textit{Hybrid-Joint} carries no such guarantee: no positivity constraint or clipping is applied to $c$ during training, so a negative, locally backscattering coefficient field was in principle reachable. That the network nonetheless converges to $c\ge0$ everywhere is therefore a property of the training outcome, not one enforced by the eddy-viscosity assumption; it is consistent with a training objective that constrains one-point statistics, such as the mean and r.m.s.\ velocity profiles, rather than the local SGS energy flux -- nothing in the loss penalizes or rewards backscatter directly. Representing backscatter would require a more general SGS-stress form than the eddy-viscosity assumption, additional constraints tied to the local energy flux, or removing the wall-parallel averaging; each would relax the structure that presently keeps training stable, and is left to future work.

\subsubsection{Neural Wall model}

\begin{figure}[htp!]
    \centerline{\includegraphics[width=\textwidth]{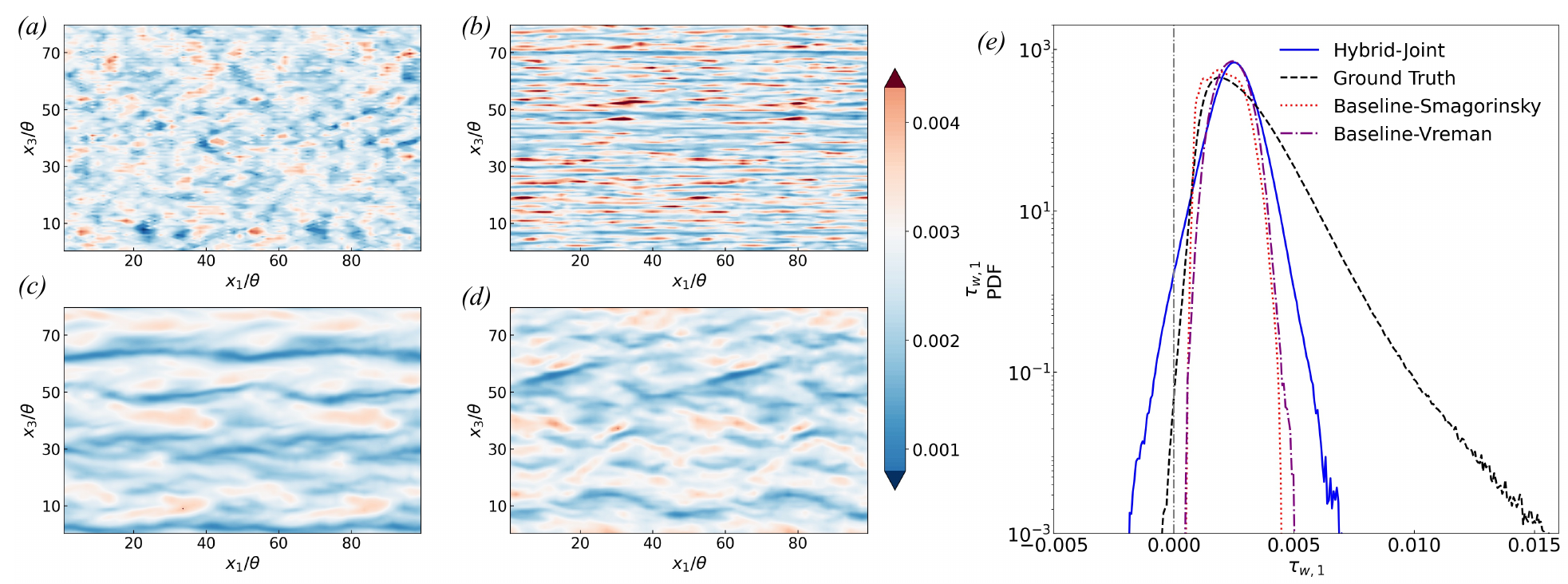}}
    \caption{Instantaneous streamwise wall-shear stress $\tau_{w,1}$ predicted by different models for Case 1 ($Re_{\theta}=600$). (a) Hybrid-Joint model. (b) Ground truth obtained by filtering the DNS solution and downsampling it to the same coarse mesh. (c) Baseline-Smagorinsky. (d) Baseline-Vreman. (e) PDF of streamwise wall-shear stress.}
    \label{fig:wall_shearstress_600}
\end{figure}

The neural wall model uses a universal skin-friction scaling law to prescribe the mean wall shear stress, while the fluctuating component is provided by the trained network. Figure~\ref{fig:wall_shearstress_600} compares the instantaneous streamwise wall-shear stress $\tau_{w,1}$ predicted for Case~1. \textit{Hybrid-Joint} gives a wall-stress field whose magnitude and spatial distribution are closer to the filtered DNS reference, though the coherent near-wall streaks are not fully captured. Both baselines, by contrast, produce clear streaky structures, but with artificially inflated length scales (figure~\ref{fig:wall_shearstress_600}a--d); this is consistent with the coarse mesh limiting what an algebraic wall-stress relation can resolve, while the network is able to represent finer spatial variation on the same grid, as also seen in the neural SGS closure (\S\,\ref{sec:sgs_analysis}).

The PDFs in figure~\ref{fig:wall_shearstress_600}(e) further highlight this difference. The reference wall shear stress is predominantly positive but has a small negative tail. \textit{Hybrid-Joint} reproduces a much broader distribution, including moderate negative fluctuations, while keeping a mean closer to the reference. \textit{Baseline-Smagorinsky} and \textit{Baseline-Vreman}, despite using different SGS closures, produce PDFs that are narrow and nearly indistinguishable from one another and admit no negative wall-shear events. This near-identity of the two baselines exposes a structural limitation of the equilibrium wall model rather than showing that wall-shear statistics are insensitive to the SGS closure: it is a fixed, instantaneous algebraic map from the resolved velocity at the matching location to $\tau_w$, with no degrees of freedom to respond to how the SGS closure shapes that resolved field. Whatever differences the SGS closure introduces upstream are therefore not carried through to the wall-stress statistics. The neural wall closure in \textit{Hybrid-Joint} has no such restriction: its correction is learned directly from the instantaneous resolved field via statistics, allowing it to reflect the coupled response of the SGS and wall closures.

\begin{figure}[htp!]
    \centerline{\includegraphics[width=\textwidth]{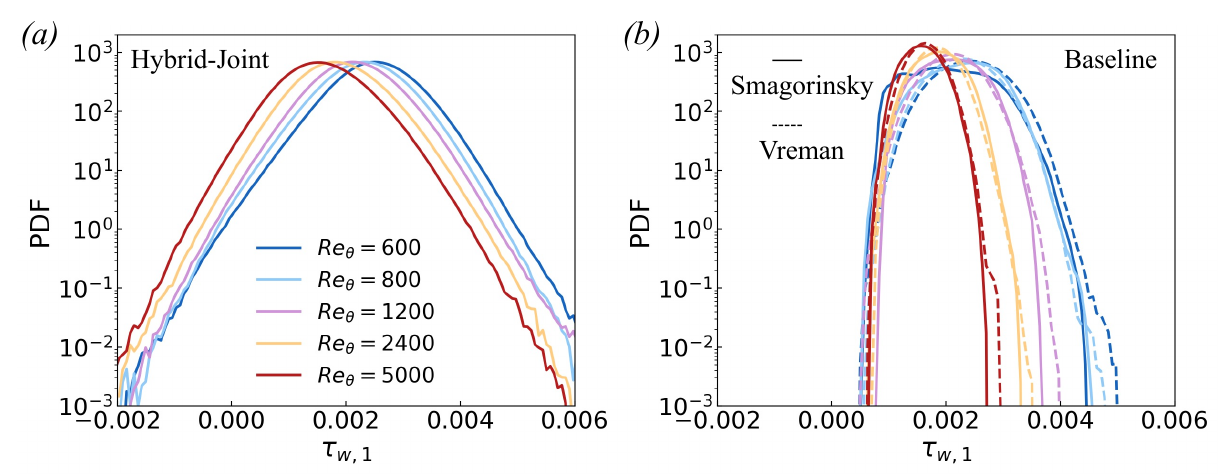}}
    \caption{PDF of the instantaneous streamwise wall-shear stress $\tau_{w,1}$ across all Reynolds numbers tested. (a) \textit{Hybrid-Joint}. (b) Two WMLES baselines, \textit{Smagorinsky} (solid) and \textit{Vreman} (dashed), which are nearly indistinguishable at every $Re_\theta$.}
    \label{fig:wall_shearstress_pdf}
\end{figure}

This pattern holds over the full Reynolds-number range tested (figure~\ref{fig:wall_shearstress_pdf}). \textit{Hybrid-Joint} clearly distinguishes the mean of $\tau_{w,1}$ across $Re_\theta$, decreasing with increasing Reynolds number as expected, and retains a broad distribution at every $Re_\theta$. Both baselines also predict the mean wall shear stress reasonably well, but their PDFs remain narrow and close to one another at every Reynolds number, for the same reason identified above: the equilibrium relation, evaluated at the matching location, leaves little room for the wall-stress statistics to depend on the SGS closure.

\begin{figure}
    \centerline{\includegraphics[width=\textwidth]{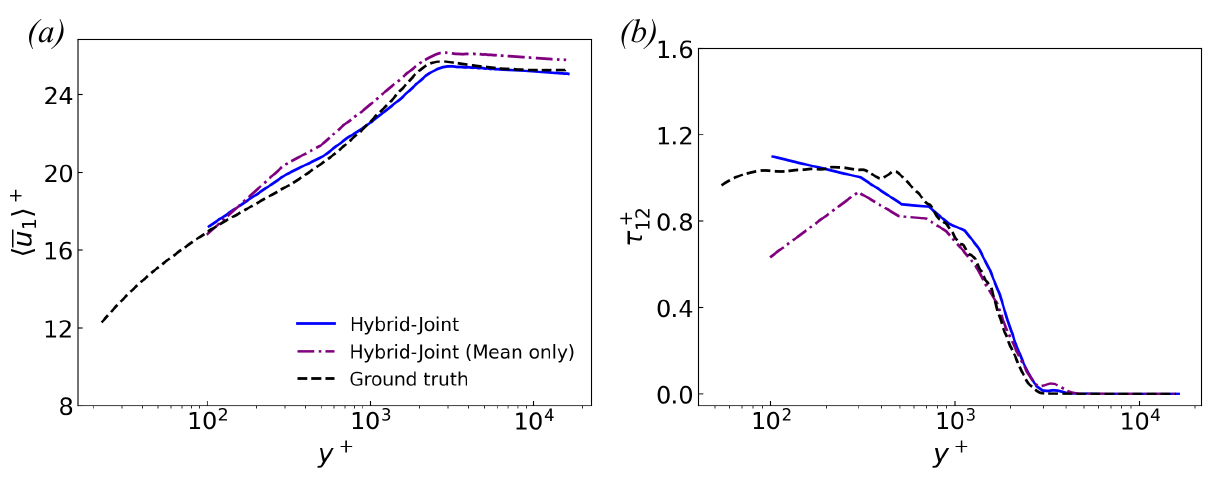}}
    \caption{Effect of the wall-stress fluctuation, evaluated at $Re_\theta = 5000$ (Case~8). \textit{Hybrid-Joint (Mean only)} retains the scaling-law mean wall stress but removes the network-predicted stress fluctuation. (a) Mean velocity profile. (b) Reynolds shear stress $\tau_{12}$.}
    \label{fig:mean_wall_effect}
\end{figure}

To isolate the contribution of the fluctuating wall-stress component, we perform an ablation in which the network correction is removed and only the scaling-law mean wall shear stress is retained. Figure~\ref{fig:mean_wall_effect} shows the results at $Re_\theta = 5000$ (Case~8). The mean-only model gives a reasonable, slightly overpredicted mean velocity profile, since the imposed mean wall stress still provides the correct inner scaling. However, the resolved Reynolds shear stress $\tau_{12}$ deteriorates markedly in the near-wall and logarithmic regions. Matching the mean wall stress alone is therefore not sufficient to recover the near-wall momentum exchange, and the wall-stress fluctuations provide unsteady forcing that the mean-only closure cannot supply. Similar ideas underlie hybrid RANS--LES approaches to WMLES, where artificial or stochastic forcing is introduced near the inner--outer layer interface to promote the development of resolved stresses~\citep{piomelli2003inner, temmerman2005hybrid, tessicini2006approximate, piomelli2008wall}. Here, the learned wall-stress fluctuations play an analogous role. Removing them narrows the $\tau_{w,1}$ distribution back toward that of the two baselines (figures~\ref{fig:wall_shearstress_600} and~\ref{fig:wall_shearstress_pdf}) and simultaneously degrades the resolved-flow statistics reported throughout \S\,\ref{sec:results}, showing that the broader distribution is not incidental but is what drives the improvement.

\subsection{Ablation study of individual components in the \textit{Hybrid-Joint} model}
\label{sec:ablation_study}

In \S\,\ref{sec:results}, we showed that \textit{Hybrid-Joint}, trained end-to-end using only turbulence statistics, gives accurate and robust \textit{a posteriori} predictions. This performance could in principle result from the design of either closure alone, from their joint optimization, or from some combination of the two. To isolate these contributions, we train two ablated variants with the same protocol and loss as the full model, and evaluate them on the most demanding extrapolation case, Case~8 at $Re_\theta = 5000$.

\textit{Hybrid-WallOnly} retains the neural wall model of \S\,\ref{sec:meth_ours} but replaces the neural SGS closure with the standard SGS model (Smagorinsky). \textit{Hybrid-SGSOnly} retains the neural SGS closure but replaces the neural wall model with the conventional equilibrium wall-stress model used in baselines. In each case, only one closure is learned and optimized with solver in the loop; the other is fixed to its conventional form. All the other training settings such as loss functions, protocol, optimization hyperparameters, etc., remain the same. Once trained, the models are evaluated in \textit{a posteriori} test for Case~8 at $Re_{\theta} = 5000$.

\begin{figure}[t]
    \centerline{\includegraphics[width=\textwidth]{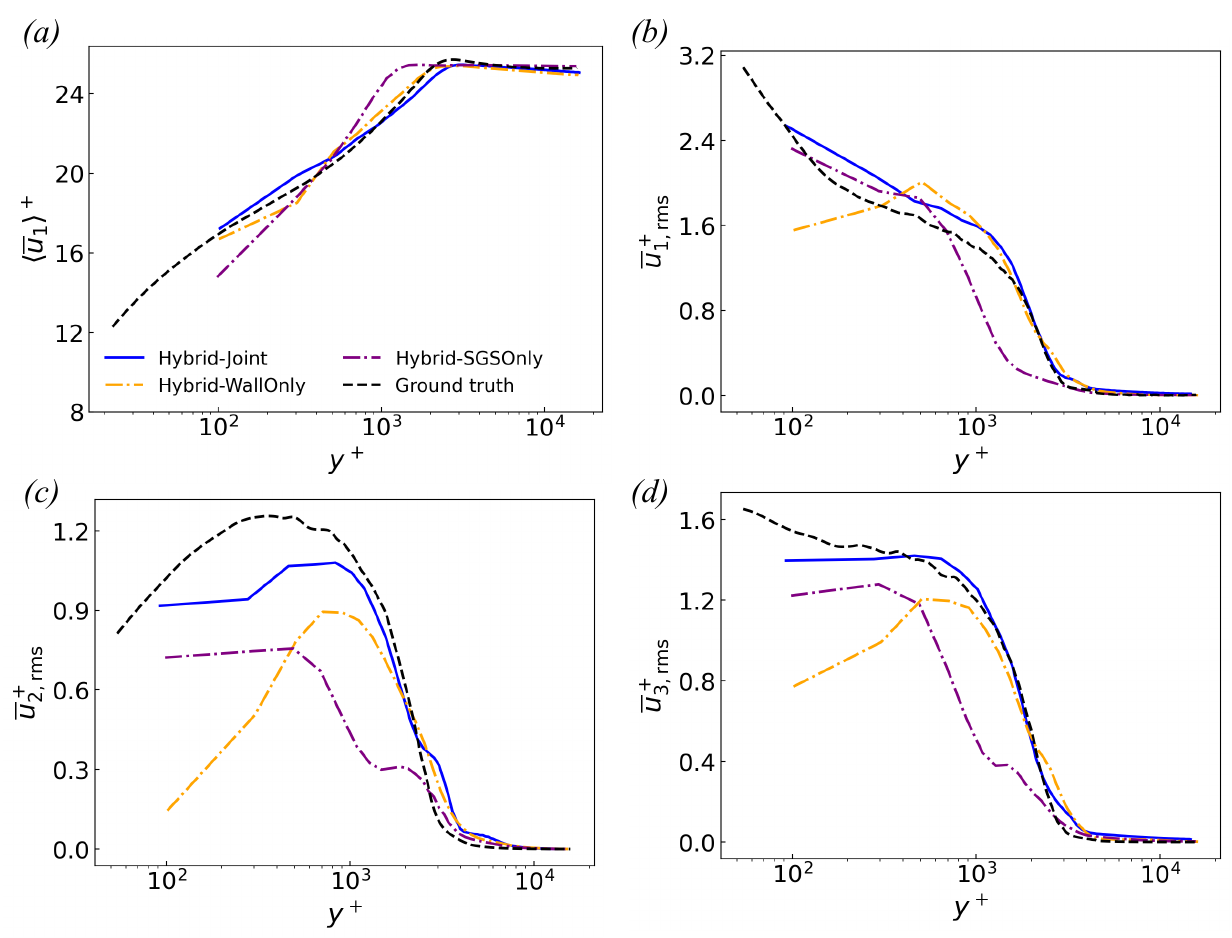}}
    \caption{Turbulence statistics for case 8 ($Re_{\theta}=5000$) under different ablation settings. (a) Mean streamwise velocity profile. (b)--(d) Root-mean-square velocity fluctuations $\overline{u}_{i,\mathrm{rms}}^{+}$.}
    \label{fig:Re5000_stat_ablation}
\end{figure}

Figure~\ref{fig:Re5000_stat_ablation} shows the results. Only the full \textit{Hybrid-Joint} model accurately predicts all the reported statistics, and the two ablated variants fail in distinct and complementary ways. \textit{Hybrid-WallOnly} gives a reasonable mean velocity profile (panel~a), consistent with its neural wall model still supplying the correct inner scaling, but it substantially underpredicts the r.m.s.\ velocity fluctuations, most severely the wall-normal component (panel~c): with the SGS closure fixed and not co-optimized with the wall model, the resolved turbulent mixing is misrepresented even though the mean profile appears adequate. \textit{Hybrid-SGSOnly} shows the opposite pattern: it fails to recover the logarithmic region of the mean velocity profile (panel~a), and the outer-layer turbulent kinetic energy is strongly underestimated (panels~b--d), even though its neural SGS closure is identical in form to that of \textit{Hybrid-Joint}. With the wall model fixed to the conventional equilibrium relation, the boundary-layer development itself is compromised, regardless of how well the interior SGS closure performs.

Neither ablation reproduces the accuracy of the full model, and each fails on a different set of quantities: \textit{Hybrid-WallOnly} predicts approximately correct mean profile but the turbulent fluctuations are incorrect, while \textit{Hybrid-SGSOnly} gets neither right. This complementary failure pattern is direct evidence that the SGS closure, the wall model and the discretization are coupled through the resolved field, as emphasized in previous studies~\citep{piomelli2002wall, rezaeiravesh2019systematic, zhang2026differentiable}: learning one closure well is not sufficient when the other remains fixed, and only the joint optimization of both within the differentiable solver recovers the full set of statistics.

\subsection{Effect of rollout length and gradient-tracking window}
\label{sec:rollout_study}

In the differentiable hybrid neural--CFD modelling framework, training stability is strongly affected by the length of the window over which gradients are back-propagated through the flow solver. The optimization involves many coupled numerical operations and unsteady flow dynamics, and this issue is particularly acute for turbulent flows, where small perturbations grow rapidly in both the forward and adjoint dynamics. The choice of tracking window is, for instance, often set by the Lyapunov time scale in data-assimilation studies~\citep{wang2022observable,wang2014least}, and similar sensitivities have been reported in differentiable neural solvers~\citep{list2022learned,list2025differentiability}; some studies enable or disable gradients through selected operators, such as the advection term or the pressure solver, to balance stability, accuracy and efficiency~\citep{franz2025pict}. The present framework requires none of these tricks, but the choice of window length still matters, which we examine here.

As described in \S\,\ref{sec:training}, the default setup uses a warm-up of $N_1 = 400$ steps ($T_1 = 1$ flow-through time), with gradients detached, followed by a gradient-tracking window of $N_2 = 100$ steps ($T_2 = 0.25$ flow-through times). The warm-up reduces the dependence on the initial condition, while the tracking window sets the statistics entering the loss. Here, we examine the sensitivity of training to these two choices.

\begin{figure}[h]
    \centerline{\includegraphics[width=\textwidth]{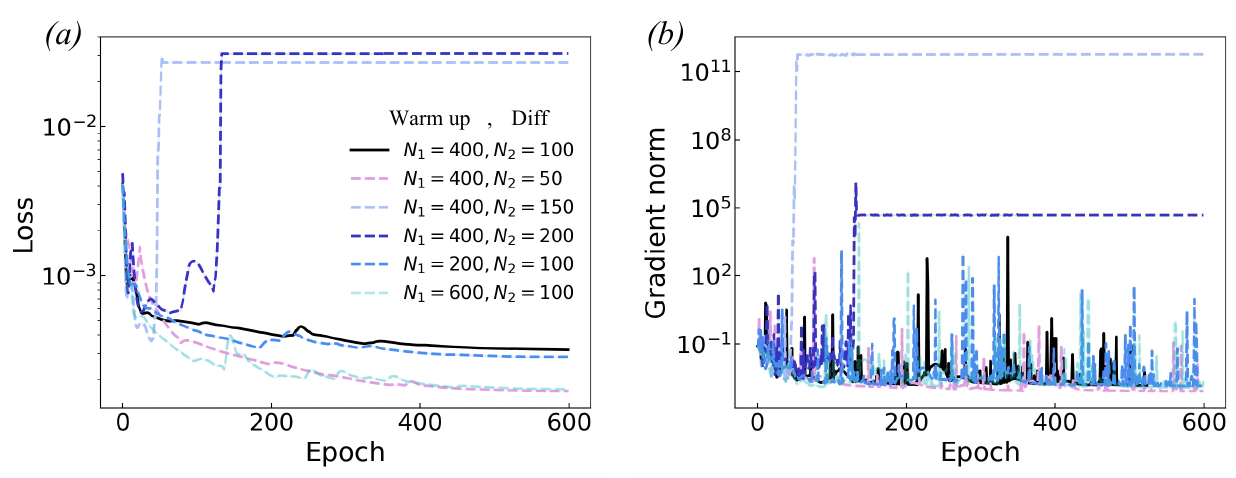}}
\caption{Stability of the Hybrid-Joint model during end-to-end differentiable training. (a) Training loss. (b) $L_2$ norm of the gradients of the trainable parameters.}
    \label{fig:training_stability}
\end{figure}

Figure~\ref{fig:training_stability} shows the training histories for several choices of $N_1$ and $N_2$. Increasing the tracking window to $N_2 = 150$ or $200$ leads to unstable training in both cases: the gradient norm eventually grows by several orders of magnitude and the loss plateaus at an elevated level instead of continuing to decrease. This non-monotonic sensitivity to the window length is itself indicative of the difficulty of controlling gradients through chaotic rollouts. Reducing the window to $N_2 = 50$ remains stable throughout training, though the statistics collected over such a short interval may be less converged; this is acceptable for the present statistically stationary flow, but a longer window is likely needed for non-equilibrium or transitional flows, chosen according to the relevant flow time scales.

The warm-up length has a weaker effect on stability. Both $N_1 = 200$ and $N_1 = 600$ (with $N_2 = 100$ fixed) train stably and reach a final loss comparable to the default $N_1 = 400$, consistent with training being initialized from fully developed fields, which already limits the influence of the initial condition on the statistics entering the loss. Increasing $N_1$ therefore brings limited benefit while increasing training cost. For simulations initialized from a mean profile or a developing state, however, a longer warm-up, typically several flow-through times, would be needed to reach a statistically developed state before the loss is evaluated.

\subsection{Computational cost and amortization}
\label{sec:cost}

Training the \textit{Hybrid-Joint} closures is a one-time cost: once trained, the same closures are reused, without further training, across new Reynolds numbers, domains, meshes and initial conditions, as demonstrated throughout \S\,\ref{sec:results}. This is a fundamental difference from conventional adjoint-based \textit{a posteriori} optimization, where the adjoint solver must be re-derived and re-optimized for each new case; the present training cost, by contrast, is amortized over every subsequent deployment of the trained model, so the relevant comparison is not training cost against a single simulation, but training cost against the savings that accrue from reuse.

Table~\ref{tab:cost} summarizes the costs involved, for Case~1 at $Re_\theta = 600$, at three distinct levels. First, a representative wall-resolved LES, with a mesh of $125\times320\times512$ and $\Delta t=0.05$, requires only 0.92 GPU-hours on our GPU-native, fully vectorised solver \texttt{Diff-FlowFSI}~\citep{fan2026diff}. This is more than $30\times$ faster than a CPU-based (OpenFOAM) solver running on 32 CPU cores, as illustrated in~\citet{fan2026diff}. The reduction reflects the efficiency of the underlying solver implementation alone, independent of mesh resolution or any embedded closure model.
Second, moving from this fully resolved WRLES mesh to the coarse WMLES mesh and time step used throughout this work, on the same GPU-native solver, further reduces the cost to $0.05$ GPU-hours per flow-through time for \textit{Hybrid-Joint}, comparable to the $0.07$ and $0.08$ GPU-hours required by \textit{Baseline-Smagorinsky} and \textit{Baseline-Vreman}, respectively -- indicating that the neural closures add negligible cost to the underlying coarse-mesh solver at inference. Third, the $40$ GPU-hours required to train \textit{Hybrid-Joint} is incurred only once and is then amortized over every subsequent inference run, at any Reynolds number or configuration within the range demonstrated in \S\,\ref{sec:results}. Taken together, these three factors, i.e., vectorized solver implementation, mesh resolution, and training amortization, compound to bring the cost of a single flow-through time from 27 CPU-hours of conventional WRLES down to $0.05$ GPU-hours of \textit{Hybrid-Joint} inference, a reduction of roughly approximately 99.8\%.

This amortization also sets the present approach apart from methods that obtain gradients through the adjoint equations, where training cost cannot be recovered in the same way because the adjoint must be recomputed for each new configuration. Reported costs illustrate the resulting gap, though comparisons across hardware and implementations are not direct and should be read as indicative rather than as an exact benchmark: \citet{mons2021ensemble} reported $9198$ CPU-hours for adjoint-based SGS optimization of a single case, against $219$ CPU-hours for a comparably accurate LES using a conventional dynamic SGS model. The cost of adjoint-based gradients also limits the length of the trajectories used to train an embedded neural closure: \citet{Kim2023GeneralizableDT} trained a graph-neural-network SGS closure using a discrete adjoint for the PDE solver, restricting training to short trajectories of $4$ consecutive snapshots in their 3D case. Among differentiable-solver approaches that instead obtain gradients natively through automatic differentiation, the gradient-tracking window has still been split into short segments of ten steps to control memory, cost and stability~\citep{zhang2026differentiable}, whereas \textit{Hybrid-Joint} trains stably over $N_2 = 100$ steps, an order of magnitude longer than both. Taken together, these figures indicate that statistics-based, end-to-end training of jointly optimized SGS and wall closures is not only accurate and robust but economically practical, with a one-time training cost amortized over every subsequent deployment.

\begin{table}
    \centering
    \begin{tabular}{lcc}
        \hline
        Item & Solver / hardware & Cost \\
        \hline
        WRLES &  OpenFOAM (32 CPU) & 27 CPU-hours \\
        WRLES & \texttt{Diff-FlowFSI} (1 GPU) & 0.92 GPU-hours \\
        \textit{Hybrid-Joint} & \texttt{Diff-FlowFSI} (1 GPU) & 0.05 GPU-hours \\
        WMLES Smagorinsky & \texttt{Diff-FlowFSI} (1 GPU) & 0.07 GPU-hours \\
        WMLES Vreman & \texttt{Diff-FlowFSI} (1 GPU) & 0.08 GPU-hours \\
        \hline
    \end{tabular}
    \caption{Inference cost for Case~1 at $Re_\theta = 600$, per flow-through time. The WRLES in OpenFOAM is reported in CPU-hours. All cases performed with \texttt{Diff-FlowFSI}, are reported in GPU-hours. Inference is performed in double precision on one NVIDIA A100 GPU. Costs across CPU and GPU hardware are not directly comparable and are reported for indicative purposes only.}
    \label{tab:cost}
\end{table}

\section{Conclusion}
\label{sec:conclusion}
In this work, we have developed a differentiable hybrid neural--CFD framework for wall-bounded turbulence at coarse space--time resolution, in which the neural SGS and wall closures are learned jointly within the discretized flow solver rather than being fitted as independent components. Each closure is constructed as a composed neural operator: a trainable network followed by a fixed, differentiable layer that carries the structure of its conventional counterpart. The neural SGS closure retains the Vreman eddy-viscosity form and learns only the operator: mapping from flow states to spatio-temporally varying coefficient field; the neural wall closure retains a skin-friction scaling law for the mean stress and learns a correction and fluctuation on top of it. Because every operation in this composition is differentiable, the two closures are trained end-to-end against the flow produced by the full discretized solver, using only low-order turbulence statistics as the training target, without instantaneous field matching or pointwise SGS-stress labels.

The trained \textit{Hybrid-Joint} model was evaluated through \textit{a posteriori} tests spanning nine configurations of a zero-pressure-gradient turbulent boundary layer, across a wide range of Reynolds numbers, computational domains, mesh resolutions and grid distributions. The model consistently outperforms two conventional WMLES baselines and reproduces the mean velocity profile, skin-friction coefficient, turbulence statistics and resolved-scale energy spectra with good accuracy, extrapolating well beyond the training Reynolds numbers and transferring to grids and domains absent from training.

Analysis of the learned components clarifies why this performance requires joint training rather than either closure alone. The neural net predicted SGS coefficient is, on average, close to the classical value for homogeneous isotropic turbulence, but its benefit lies in its retained spatial structure and flow-dependent fluctuations rather than in this mean, though this leaves the closure unable to represent backscatter, a limitation shared by constant eddy-viscosity-based models. At the wall, the conventional equilibrium relation is structurally unable to reflect how the SGS closure shapes the resolved flow, whereas the learned wall-stress fluctuations do, and are shown by ablation to be necessary for the near-wall Reynolds shear stress. A further ablation, learning either closure alone while fixing the other, shows that the two fail in distinct and complementary ways: only their joint optimization recovers the full set of statistics. Together, these results confirm that the SGS closure, the wall closure and the numerical discretization are coupled through the resolved field, and that this coupling must be addressed in training rather than in the choice of closure form alone.

The framework is also computationally practical. Once trained, at a one-time cost, the same closures are reused without retraining across every Reynolds number, domain, mesh and initial condition examined, so training cost is amortized over repeated deployment rather than incurred once per case, unlike adjoint-based \textit{a posteriori} optimization. Stable training over a $100$-step gradient-tracking window is achieved from scratch, without the short windows, pretraining or curriculum strategies that other differentiable-solver studies have required, which we attribute to the robustness of our in-house differentiable CFD platform (\texttt{Diff-FlowFSI}) and to the structure-constrained closures, whose learnable outputs remain confined to a scalar coefficient field and a stress correction.

The present framework thus provides a route to data-informed WMLES closures with improved \textit{a posteriori} accuracy and generalizability, achieved by learning the coupling between the SGS closure, the wall closure and the discretization directly, rather than by improving either closure in isolation. Its present scope is limited to the equilibrium, statistically stationary zero-pressure-gradient boundary layer; extending it to non-equilibrium flows, including adverse-pressure-gradient, separated and transitional boundary layers, is left to future work. Representing backscatter would require relaxing the eddy-viscosity form of the SGS closure, and extending the training objective to include energy transfer, spectral content or multi-point statistics is a natural next step, provided the stability of end-to-end differentiable training is preserved.

\section*{Acknowledgment}
 The authors would like to acknowledge the funds from Office of Naval Research under award number N00014-23-1-2071, National Institutes of Health under award number 1R01HL177814, and National Science Foundation under award number OAC-2047127.
\section*{Compliance with Ethical Standards}
Conflict of Interest: The authors declare that they have no conflict of interest.

\clearpage

\appendix
\begin{appen}

\section{Neural network structures}
\label{sec:neural_net}

The neural-network architectures and training parameters are summarized in 
table~\ref{tab:network_parameters}.

\begin{table}
    \centering
    \begin{tabular}{lll}
        \hline
        Module & Parameter & Value \\
        \hline
        SGS model   & Architecture          & 3-D U-Net \\
                    & Channel progression   & $[32, 16, 8, 4, 1]$ \\
                    & Convolution kernel    & $3 \times 3 \times 3$ \\
                    & Input                 & Filtered velocity, pressure and strain-rate tensor \\
                    & Output                & Vreman coefficient field, $c_m$ \\
        \hline
        Wall model  & Architecture          & 2-D U-Net \\
                    & Channel progression   & $[64, 32, 32, 8, 2]$ \\
                    & Convolution kernel    & $3 \times 3$ \\
                    & Input                 & Filtered velocity and pressure at the third off-wall cell \\
                    & Output                & Wall-stress correction and fluctuations, $\tau'_{w,1}$, $\tau'_{w,3}$ \\
        \hline
        Training    & Optimiser             & Adam (\texttt{Optax}) \\
                    & Learning-rate schedule & Cosine decay \\
                    & Initial learning rate  & $1 \times 10^{-4}$ \\
                    & Total epochs           & $600$ \\
                    & Gradient clipping      & $L_2$-norm bound of $10$ \\
                    & Spatial patch size     & $64 \times 64$ \\
                    & Warm-up window         & $T_1 = 1$ ($N_1 = 400$ steps) \\
                    & Gradient-tracking window & $T_2 = 0.25$ ($N_2 = 100$ steps) \\
                    & Training $Re_{\theta}$ & $600$, $1200$ \\
                    & Loss weighting         & Inverse-magnitude, $w_k = 1/|\mathcal{L}_k|$ \\
        \hline
    \end{tabular}
    \caption{Neural-network architectures and training parameters of the proposed hybrid framework. The SGS closure uses a 3-D U-Net that outputs the Vreman coefficient field $c_m$; the wall closure uses a 2-D U-Net that outputs the two wall-parallel stress fluctuations.}
    \label{tab:network_parameters}
\end{table}

\section{Baselines}
\label{sec:baseline}

For \textit{Baseline-Smagorinsky}, we employ the constant-coefficient Smagorinsky model as the SGS closure. Following the general eddy-viscosity form in equation~\eqref{eq:generic_eddyvisc}, the model is written as
\begin{equation}
\mathcal{F}(\overline{u}_i)=|\overline{S}|,
\qquad
c=f_v (C_s\Delta)^2,
\label{eq:constant_sgs}
\end{equation}
where $C_s=0.2$ is the Smagorinsky coefficient, $\Delta=(\Delta_1\Delta_2\Delta_3)^{1/3}$ is the filter width based on the local grid spacing, and
\begin{equation}
|\overline{S}|=(2\overline{S}_{ij}\overline{S}_{ij})^{1/2}
\end{equation}
is the magnitude of the resolved strain-rate tensor. The factor $f_v$ is a near-wall damping function.

The constant Smagorinsky model does not, by itself, recover the correct near-wall asymptotic behaviour of the SGS stress. A damping function is therefore required, particularly in the near-wall region, here taken as $y^+<100$. The classical choice is the van Driest damping function~\citep{van1956turbulent},
\begin{equation}
f_v=
\left[
1-\exp\left(-\frac{y^+}{A^+}\right)
\right]^2,
\end{equation}
where $y^+$ is the wall-normal distance in wall units and $A^+=25$. In the present baseline WMLES, however, we instead employ the third-order damping form
\begin{equation}
f_v=
1-\exp\left[
-\left(\frac{y^+}{A^+}\right)^3
\right].
\end{equation}
This choice yields $f_v\sim (y^+)^3$ as $y^+\rightarrow 0$, consistent with the expected near-wall scaling of the SGS shear stress, $\tau_{12}\sim (y^+)^3$. Consistent with previous studies~\citep{piomelli1988models,nabae2025large}, our preliminary tests further indicate that this third-order damping form provides more accurate prediction of the skin-friction coefficient than the classical van Driest damping function for the baseline cases considered here.

At the wall, the equilibrium wall-stress model is employed. Following equation~\eqref{eq:generic_wallmodel}, the function $f$ is taken to be the Spalding wall function,
\begin{equation}
u^{+}=y^{+}-e^{-\kappa B}\left[e^{\kappa u^{+}}-1-\kappa u^{+}-\frac{1}{2}(\kappa u^{+})^{2}-\frac{1}{6}(\kappa u^{+})^{3}\right],
\label{eq:wall_model}
\end{equation}
where $\kappa=0.4$ and $B=5.0$, with $u^{+}=\overline{U}/u_{\tau}$ and $y^{+}=y_m u_{\tau}/\nu$. Here, $\overline{U}$ denotes the magnitude of the filtered wall-parallel velocity,
\begin{equation}
\overline{U}=\sqrt{\overline{u}_1^2+\overline{u}_3^2},
\label{eq:wall_parallel_vel}
\end{equation}
$y_m$ is the wall distance at the matching location, taken here as the third off-wall cell centre, and $u_{\tau}$ is the friction velocity, obtained iteratively using Newton's method. The resulting $u_{\tau}$ is then substituted into equation~\eqref{eq:modified_nu} to determine the wall eddy viscosity.

For \textit{Baseline-Vreman}, the SGS closure is given by the Vreman model, where $\mathcal{F}(\overline{u}_i)$ is defined in equations~\eqref{eq:vreman_form} and \eqref{eq:vreman_Bbeta}. In the original constant-coefficient Vreman model, the coefficient is commonly prescribed as $c=2.5C_s^2$, where $C_s$ is the Smagorinsky constant~\citep{vreman2004eddy}. In the present baseline, we set $c=0.07$. Although other values have been used in the literature, this standard choice is adopted here for illustration. The same equilibrium wall-stress model, defined by equation~\eqref{eq:wall_model}, is again employed for consistency.

\section{Additional results}
\label{sec:additional_results}

\begin{figure}[h]
    \centerline{\includegraphics[width=0.9\textwidth]{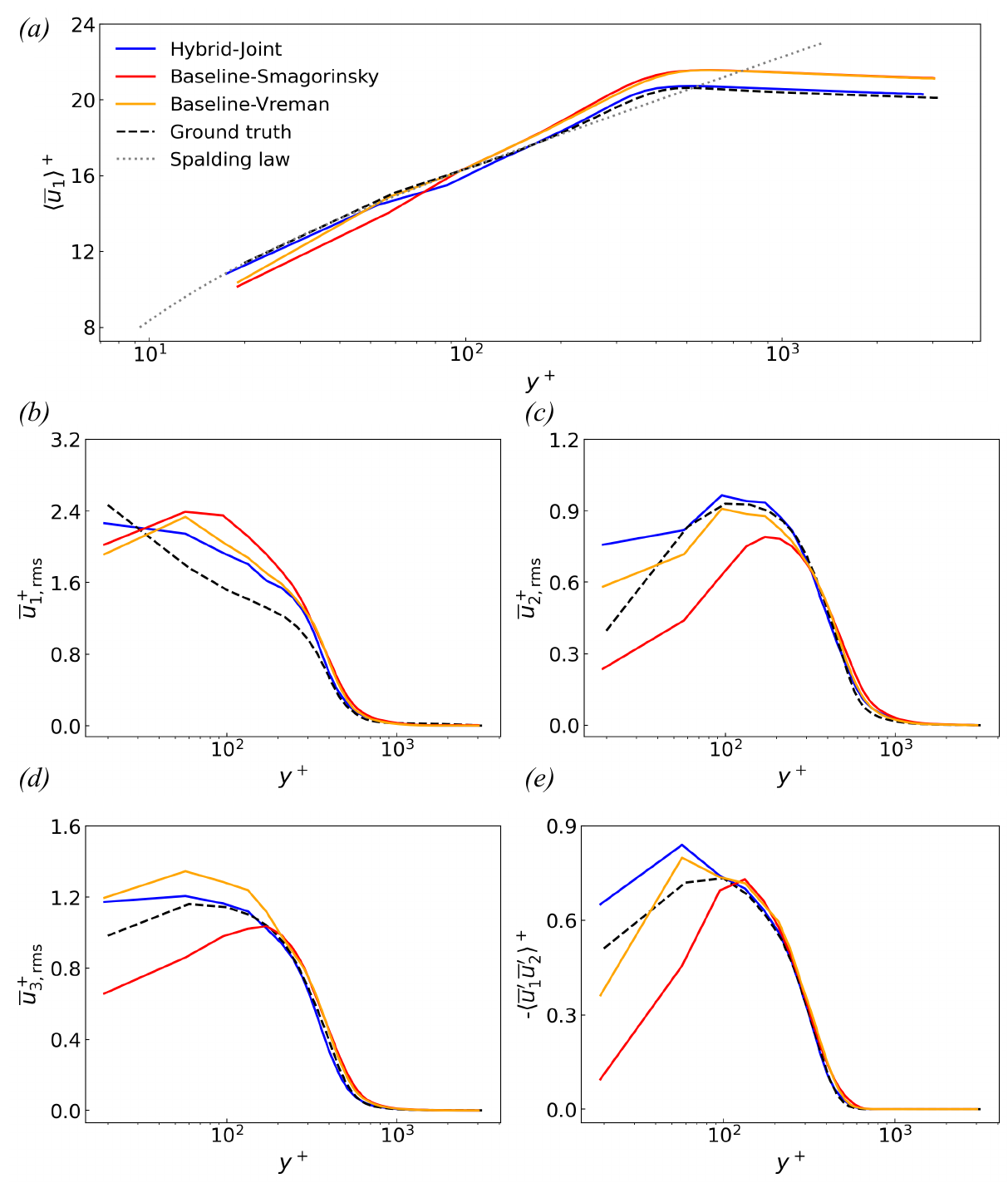}}
    \caption{Turbulence statistics for Case 2 ($Re_{\theta}=800$), compared with the two baselines and the ground truth (DNS) (a) Mean streamwise velocity profile. (b)--(d) Root-mean-square (RMS) velocity fluctuations $\overline{u}_{i,\mathrm{rms}}^{+}$. (e) Root-mean-square Reynolds shear stress $-\langle\overline{u}_1'\overline{u}_2'\rangle^{+}$. All statistics are normalized by the friction velocity $u_{\tau}$.}
    \label{fig:Re800_stat}
\end{figure}

\begin{figure}[h]
    \centerline{\includegraphics[width=0.9\textwidth]{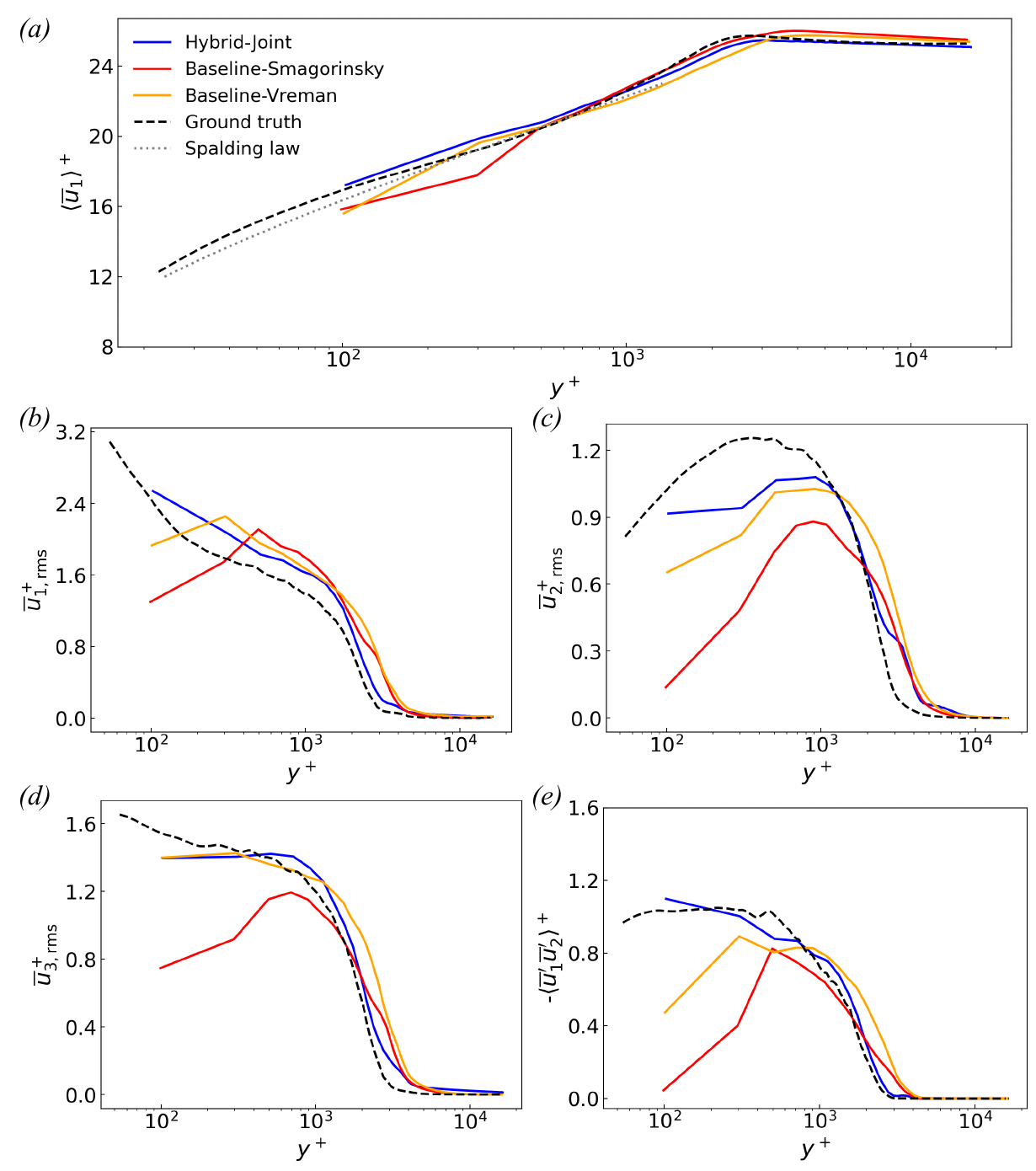}}
    \caption{Turbulence statistics for Case 8 ($Re_{\theta}=5000$), compared with the two baselines and the ground truth (wall-resolved-quality fine-mesh WMLES) (a) Mean streamwise velocity profile. (b)--(d) Root-mean-square (RMS) velocity fluctuations $\overline{u}_{i,\mathrm{rms}}^{+}$. (e) Root-mean-square Reynolds shear stress $-\langle\overline{u}_1'\overline{u}_2'\rangle^{+}$. All statistics are normalized by the friction velocity $u_{\tau}$.}
    \label{fig:Re5000_stat}
\end{figure}

\begin{figure}[h]
    \centerline{\includegraphics[width=\textwidth]{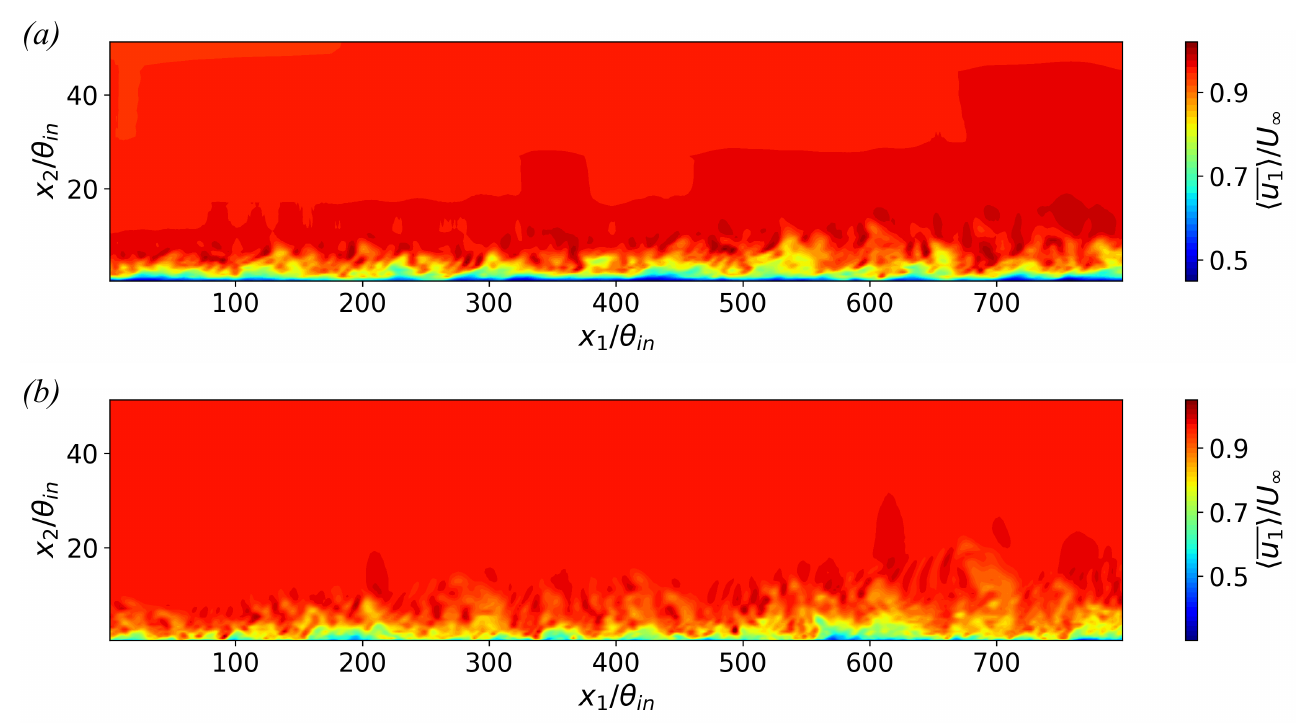}}
    \caption{Instantaneous streamwise velocity contour for Case 9 from two baselines ($Re_{\theta,\mathrm{in}}=5000$): (a) Baseline-Smagorinsky. (b) Baseline-Vreman.}
    \label{fig:Re5000_long_classic_contour}
\end{figure}

\begin{figure}[h]
    \centerline{\includegraphics[width=\textwidth]{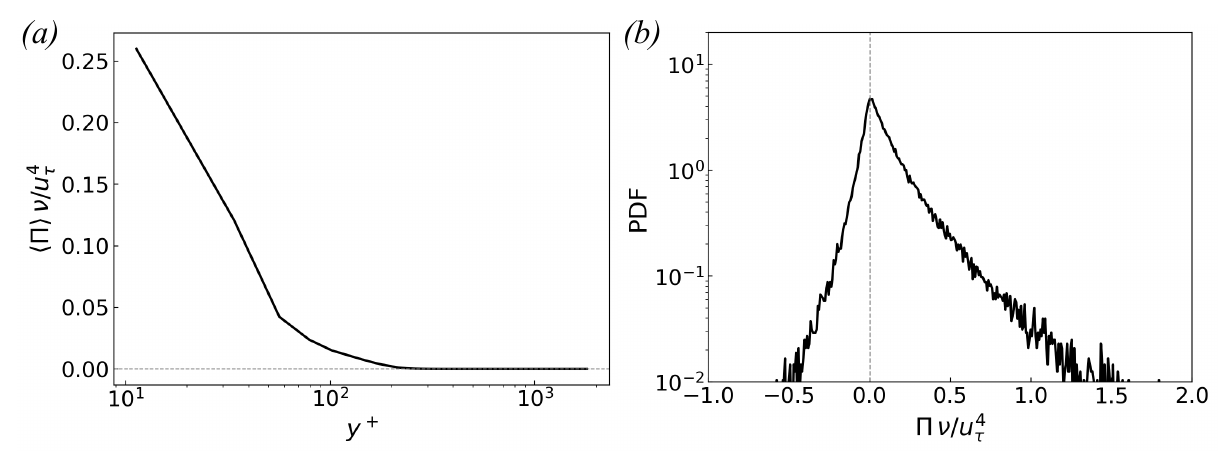}}
    \caption{Energy transfer between the resolved and subgrid scales for \textit{a priori} test at $Re_{\theta} = 600$. (a) Mean SGS energy-transfer rate $\langle\Pi\rangle$ in the wall-normal direction. (b) Probability
density function (PDF) of the instantaneous energy-transfer rate $\Pi$ at $x_2 / \delta_{99} = 0.2$.}
\label{fig:Re600_priori_enenrgy}
\end{figure}

\end{appen}\clearpage

\bibliographystyle{jfm}
\bibliography{v0/otherRef,v0/myRef}

\begin{thebibliography}{73}
\expandafter\ifx\csname natexlab\endcsname\relax\def\natexlab#1{#1}\fi
\def\au#1{#1} \def\ed#1{#1} \def\yr#1{#1}\def\at#1{#1}\def\jt#1{\textit{#1}} \def\bt#1{#1}\def\bvol#1{\textbf{#1}} \def\vol#1{#1} \def\pg#1{#1} \def\publ#1{#1}\def\arxiv#1{#1}\def\org#1{#1}\def\st#1{\textit{#1}}

\bibitem[Abkar \& Port{\'e}-Agel(2015)]{abkar2015influence}
{\sc \au{Abkar, Mahdi} \& \au{Port{\'e}-Agel, Fernando}} \yr{2015}  \at{Influence of atmospheric stability on wind-turbine wakes: A large-eddy simulation study}.  \jt{Physics of Fluids}  \bvol{27}~(3).

\bibitem[Akhare {\em et~al.\/}(2025)Akhare, Luo \& Wang]{akhare2025hybridndiff}
{\sc \au{Akhare, Deepak}, \au{Luo, Tengfei} \& \au{Wang, Jian-Xun}} \yr{2025}  \at{{HybridNDiff-UQ}: Uncertainty quantification for hybrid neural differentiable modeling}.  \jt{Theoretical and Applied Mechanics Letters}  \pg{p. 100609}.

\bibitem[Bae \& Lozano-Dur{\'a}n(2021)]{bae2021effect}
{\sc \au{Bae, H~Jane} \& \au{Lozano-Dur{\'a}n, Adri{\'a}n}} \yr{2021}  \at{Effect of wall boundary conditions on a wall-modeled large-eddy simulation in a finite-difference framework}.  \jt{Fluids}  \bvol{6}~(3),  \pg{112}.

\bibitem[Bardina {\em et~al.\/}(1980)Bardina, Ferziger \& Reynolds]{bardina1980improved}
{\sc \au{Bardina, Jorge}, \au{Ferziger, J} \& \au{Reynolds, WC}} \yr{1980} Improved subgrid-scale models for large-eddy simulation.  \bt{In {\em 13th Fluid and Plasmadynamics Conference\/}},  \pg{p. 1357}.

\bibitem[Bezgin {\em et~al.\/}(2023)Bezgin, Buhendwa \& Adams]{bezgin2023jax}
{\sc \au{Bezgin, Deniz~A}, \au{Buhendwa, Aaron~B} \& \au{Adams, Nikolaus~A}} \yr{2023}  \at{Jax-fluids: A fully-differentiable high-order computational fluid dynamics solver for compressible two-phase flows}.  \jt{Computer Physics Communications}  \bvol{282},  \pg{108527}.

\bibitem[Bose \& Park(2018)]{bose2018wall}
{\sc \au{Bose, Sanjeeb~T} \& \au{Park, George~Ilhwan}} \yr{2018}  \at{Wall-modeled large-eddy simulation for complex turbulent flows}.  \jt{Annual Review of Fluid Mechanics}  \bvol{50},  \pg{535--561}.

\bibitem[Cantwell(2021)]{cantwell2021integral}
{\sc \au{Cantwell, Brian~J.}} \yr{2021}  \at{Integral measures of the zero pressure gradient boundary layer over the {R}eynolds number range {$0 \leq R_\tau < \infty$}}.  \jt{Physics of Fluids}  \bvol{33}~(8),  \pg{085108}.

\bibitem[Chapman(1979)]{chapman1979computational}
{\sc \au{Chapman, Dean~R}} \yr{1979}  \at{Computational aerodynamics development and outlook}.  \jt{AIAA Journal}  \bvol{17}~(12),  \pg{1293--1313}.

\bibitem[Chedevergne(2025)]{chedevergne2025characterisation}
{\sc \au{Chedevergne, Fran{\c{c}}ois}} \yr{2025}  \at{Characterisation and modelling of interscale energy transfers in high {R}eynolds number boundary layers}.  \jt{Journal of Fluid Mechanics}  \bvol{1024},  \pg{A28}.

\bibitem[Clark {\em et~al.\/}(1979)Clark, Ferziger \& Reynolds]{clark1979evaluation}
{\sc \au{Clark, Robert~A}, \au{Ferziger, Joel~H} \& \au{Reynolds, William~Craig}} \yr{1979}  \at{Evaluation of subgrid-scale models using an accurately simulated turbulent flow}.  \jt{Journal of Fluid Mechanics}  \bvol{91}~(1),  \pg{1--16}.

\bibitem[Du {\em et~al.\/}(2026)Du, Li, Xu \& Wang]{du2026difvm}
{\sc \au{Du, Pan}, \au{Li, Yongqi}, \au{Xu, Mingqi} \& \au{Wang, Jian-Xun}} \yr{2026}  \at{{DiFVM}: A vectorized graph-based finite volume solver for differentiable cfd on unstructured meshes}.  \jt{arXiv preprint arXiv:2603.15920} .

\bibitem[Duraisamy {\em et~al.\/}(2019)Duraisamy, Iaccarino \& Xiao]{duraisamy2019turbulence}
{\sc \au{Duraisamy, Karthik}, \au{Iaccarino, Gianluca} \& \au{Xiao, Heng}} \yr{2019}  \at{Turbulence modeling in the age of data}.  \jt{Annual Review of Fluid Mechanics}  \bvol{51}~(1),  \pg{357--377}.

\bibitem[Fan {\em et~al.\/}(2025{\natexlab{{\em a\/}}})Fan, Zhao \& Wan]{fan2025near}
{\sc \au{Fan, Guoqing}, \au{Zhao, Weiwen} \& \au{Wan, Decheng}} \yr{2025{\natexlab{{\em a\/}}}}  \at{Near-wall turbulent fluctuations and coherent structures in wall-modeled large-eddy simulation}.  \jt{Physics of Fluids}  \bvol{37}~(9).

\bibitem[Fan {\em et~al.\/}(2025{\natexlab{{\em b\/}}})Fan, Akhare \& Wang]{fan2025neural}
{\sc \au{Fan, Xiantao}, \au{Akhare, Deepak} \& \au{Wang, Jian-Xun}} \yr{2025{\natexlab{{\em b\/}}}}  \at{Neural differentiable modeling with diffusion-based super-resolution for two-dimensional spatiotemporal turbulence}.  \jt{Computer Methods in Applied Mechanics and Engineering}  \bvol{433},  \pg{117478}.

\bibitem[Fan {\em et~al.\/}(2026)Fan, Liu, Wang \& Wang]{fan2026diff}
{\sc \au{Fan, Xiantao}, \au{Liu, Xin-Yang}, \au{Wang, Meng} \& \au{Wang, Jian-Xun}} \yr{2026}  \at{{Diff-FlowFSI}: A {GPU}-optimized differentiable {CFD} platform for high-fidelity turbulence and {FSI} simulations}.  \jt{Computer Methods in Applied Mechanics and Engineering}  \bvol{448},  \pg{118455}.

\bibitem[Fan \& Wang(2024)]{fan2024differentiable}
{\sc \au{Fan, Xiantao} \& \au{Wang, Jian-Xun}} \yr{2024}  \at{Differentiable hybrid neural modeling for fluid-structure interaction}.  \jt{Journal of Computational Physics}  \bvol{496},  \pg{112584}.

\bibitem[Fernholz \& Finley(1996)]{fernholz1996incompressible}
{\sc \au{Fernholz, Hans-Hermann} \& \au{Finley, PJ}} \yr{1996}  \at{The incompressible zero-pressure-gradient turbulent boundary layer: an assessment of the data}.  \jt{Progress in Aerospace Sciences}  \bvol{32}~(4),  \pg{245--311}.

\bibitem[Fowler {\em et~al.\/}(2022)Fowler, Zaki \& Meneveau]{fowler2022lagrangian}
{\sc \au{Fowler, Mitchell}, \au{Zaki, Tamer~A} \& \au{Meneveau, Charles}} \yr{2022}  \at{A lagrangian relaxation towards equilibrium wall model for large eddy simulation}.  \jt{Journal of Fluid Mechanics}  \bvol{934},  \pg{A44}.

\bibitem[Franz {\em et~al.\/}(2025)Franz, Wei, Guastoni \& Thuerey]{franz2025pict}
{\sc \au{Franz, Aleksandra}, \au{Wei, Hao}, \au{Guastoni, Luca} \& \au{Thuerey, Nils}} \yr{2025}  \at{{PICT}--a differentiable, {GPU}-accelerated multi-block {PISO} solver for simulation-coupled learning tasks in fluid dynamics}.  \jt{Journal of Computational Physics}  \pg{p. 114433}.

\bibitem[Fukami {\em et~al.\/}(2019)Fukami, Nabae, Kawai \& Fukagata]{fukami2019synthetic}
{\sc \au{Fukami, Kai}, \au{Nabae, Yusuke}, \au{Kawai, Ken} \& \au{Fukagata, Koji}} \yr{2019}  \at{Synthetic turbulent inflow generator using machine learning}.  \jt{Physical Review Fluids}  \bvol{4}~(6),  \pg{064603}.

\bibitem[Germano {\em et~al.\/}(1991)Germano, Piomelli, Moin \& Cabot]{germano1991dynamic}
{\sc \au{Germano, Massimo}, \au{Piomelli, Ugo}, \au{Moin, Parviz} \& \au{Cabot, William~H}} \yr{1991}  \at{A dynamic subgrid-scale eddy viscosity model}.  \jt{Physics of Fluids A: Fluid dynamics}  \bvol{3}~(7),  \pg{1760--1765}.

\bibitem[Guan {\em et~al.\/}(2022)Guan, Chattopadhyay, Subel \& Hassanzadeh]{guan2022stable}
{\sc \au{Guan, Yifei}, \au{Chattopadhyay, Ashesh}, \au{Subel, Adam} \& \au{Hassanzadeh, Pedram}} \yr{2022}  \at{Stable a posteriori {LES} of {2D} turbulence using convolutional neural networks: Backscattering analysis and generalization to higher {R}e via transfer learning}.  \jt{Journal of Computational Physics}  \bvol{458},  \pg{111090}.

\bibitem[Guastoni {\em et~al.\/}(2021)Guastoni, G{\"u}emes, Ianiro, Discetti, Schlatter, Azizpour \& Vinuesa]{guastoni2021convolutional}
{\sc \au{Guastoni, Luca}, \au{G{\"u}emes, Alejandro}, \au{Ianiro, Andrea}, \au{Discetti, Stefano}, \au{Schlatter, Philipp}, \au{Azizpour, Hossein} \& \au{Vinuesa, Ricardo}} \yr{2021}  \at{Convolutional-network models to predict wall-bounded turbulence from wall quantities}.  \jt{Journal of Fluid Mechanics}  \bvol{928},  \pg{A27}.

\bibitem[Huang {\em et~al.\/}(2026)Huang, Leung \& Bae]{huang2026consistency}
{\sc \au{Huang, Xinyi}, \au{Leung, Sze~Chai} \& \au{Bae, H~Jane}} \yr{2026}  \at{Consistency requirement of data-driven subgrid-scale modeling in large-eddy simulation}.  \jt{Physical Review Fluids}  \bvol{11}~(1),  \pg{014602}.

\bibitem[Kawai \& Larsson(2012)]{kawai2012wall}
{\sc \au{Kawai, Soshi} \& \au{Larsson, Johan}} \yr{2012}  \at{Wall-modeling in large eddy simulation: Length scales, grid resolution, and accuracy}.  \jt{Physics of Fluids}  \bvol{24}~(1).

\bibitem[Kim {\em et~al.\/}(2025)Kim, Shankar, Viswanathan \& Maulik]{Kim2023GeneralizableDT}
{\sc \au{Kim, Hojin}, \au{Shankar, Varun}, \au{Viswanathan, Venkatasubramanian} \& \au{Maulik, Romit}} \yr{2025} Generalizable data-driven turbulence closure modeling on unstructured grids with differentiable physics,  \arxiv{arXiv: 2307.13533}.

\bibitem[Kochkov {\em et~al.\/}(2021)Kochkov, Smith, Alieva, Wang, Brenner \& Hoyer]{kochkov2021machine}
{\sc \au{Kochkov, Dmitrii}, \au{Smith, Jamie~A}, \au{Alieva, Ayya}, \au{Wang, Qing}, \au{Brenner, Michael~P} \& \au{Hoyer, Stephan}} \yr{2021}  \at{Machine learning--accelerated computational fluid dynamics}.  \jt{Proceedings of the National Academy of Sciences}  \bvol{118}~(21),  \pg{e2101784118}.

\bibitem[Kohl {\em et~al.\/}(2026)Kohl, Chen \& Thuerey]{kohl2026benchmarking}
{\sc \au{Kohl, Georg}, \au{Chen, Li-Wei} \& \au{Thuerey, Nils}} \yr{2026}  \at{Benchmarking autoregressive conditional diffusion models for turbulent flow simulation}.  \jt{Neural Networks}  \pg{p. 108641}.

\bibitem[Larsson {\em et~al.\/}(2016)Larsson, Kawai, Bodart \& Bermejo-Moreno]{larsson2016large}
{\sc \au{Larsson, Johan}, \au{Kawai, Soshi}, \au{Bodart, Julien} \& \au{Bermejo-Moreno, Ivan}} \yr{2016}  \at{Large eddy simulation with modeled wall-stress: recent progress and future directions}.  \jt{Mechanical Engineering Reviews}  \bvol{3}~(1),  \pg{15--00418}.

\bibitem[Li {\em et~al.\/}(2023)Li, Peng, Yuan \& Wang]{li2023long}
{\sc \au{Li, Zhijie}, \au{Peng, Wenhui}, \au{Yuan, Zelong} \& \au{Wang, Jianchun}} \yr{2023}  \at{Long-term predictions of turbulence by implicit {U}-{N}et enhanced f{o}urier neural operator}.  \jt{Physics of Fluids}  \bvol{35}~(7).

\bibitem[Lilly(1966)]{lilly1966application}
{\sc \au{Lilly, D}} \yr{1966}  \at{On the application of the eddy viscosity concept in the inertial sub-range of turbulence}.  \jt{NCAR Report} .

\bibitem[Ling \& Lozano-Duran(2025)]{ling2025numerically}
{\sc \au{Ling, Yuenong} \& \au{Lozano-Duran, Adrian}} \yr{2025} Numerically consistent data-driven subgrid-scale model via data assimilation and machine learning.  \bt{In {\em AIAA SCITECH 2025 Forum\/}},  \pg{p. 1280}.

\bibitem[List {\em et~al.\/}(2025)List, Chen, Bali \& Thuerey]{list2025differentiability}
{\sc \au{List, Bjoern}, \au{Chen, Li-Wei}, \au{Bali, Kartik} \& \au{Thuerey, Nils}} \yr{2025}  \at{Differentiability in unrolled training of neural physics simulators on transient dynamics}.  \jt{Computer Methods in Applied Mechanics and Engineering}  \bvol{433},  \pg{117441}.

\bibitem[List {\em et~al.\/}(2022)List, Chen \& Thuerey]{list2022learned}
{\sc \au{List, Bj{\"o}rn}, \au{Chen, Li-Wei} \& \au{Thuerey, Nils}} \yr{2022}  \at{Learned turbulence modelling with differentiable fluid solvers: physics-based loss functions and optimisation horizons}.  \jt{Journal of Fluid Mechanics}  \bvol{949},  \pg{A25}.

\bibitem[Lund {\em et~al.\/}(1998)Lund, Wu \& Squires]{lund1998generation}
{\sc \au{Lund, Thomas~S}, \au{Wu, Xiaohua} \& \au{Squires, Kyle~D}} \yr{1998}  \at{Generation of turbulent inflow data for spatially-developing boundary layer simulations}.  \jt{Journal of computational physics}  \bvol{140}~(2),  \pg{233--258}.

\bibitem[MacArt {\em et~al.\/}(2021)MacArt, Sirignano \& Freund]{macart2021embedded}
{\sc \au{MacArt, Jonathan~F}, \au{Sirignano, Justin} \& \au{Freund, Jonathan~B}} \yr{2021}  \at{Embedded training of neural-network subgrid-scale turbulence models}.  \jt{Physical Review Fluids}  \bvol{6}~(5),  \pg{050502}.

\bibitem[Maulik {\em et~al.\/}(2019)Maulik, San, Rasheed \& Vedula]{maulik2019subgrid}
{\sc \au{Maulik, Romit}, \au{San, Omer}, \au{Rasheed, Adil} \& \au{Vedula, Prakash}} \yr{2019}  \at{Subgrid modelling for two-dimensional turbulence using neural networks}.  \jt{Journal of Fluid Mechanics}  \bvol{858},  \pg{122--144}.

\bibitem[McConkey {\em et~al.\/}(2021)McConkey, Yee \& Lien]{mcconkey2021curated}
{\sc \au{McConkey, Ryley}, \au{Yee, Eugene} \& \au{Lien, Fue-Sang}} \yr{2021}  \at{A curated dataset for data-driven turbulence modelling}.  \jt{Scientific Data}  \bvol{8}~(1),  \pg{255}.

\bibitem[Meneveau \& Katz(2000)]{meneveau2000scale}
{\sc \au{Meneveau, Charles} \& \au{Katz, Joseph}} \yr{2000}  \at{Scale-invariance and turbulence models for large-eddy simulation}.  \jt{Annual Review of Fluid Mechanics}  \bvol{32}~(1),  \pg{1--32}.

\bibitem[Mons {\em et~al.\/}(2021)Mons, Du \& Zaki]{mons2021ensemble}
{\sc \au{Mons, Vincent}, \au{Du, Yifan} \& \au{Zaki, Tamer~A}} \yr{2021}  \at{Ensemble-variational assimilation of statistical data in large-eddy simulation}.  \jt{Physical Review Fluids}  \bvol{6}~(10),  \pg{104607}.

\bibitem[Nabae {\em et~al.\/}(2025)Nabae, Inagaki, Kobayashi, Gotoda \& Fukagata]{nabae2025large}
{\sc \au{Nabae, Yusuke}, \au{Inagaki, Kazuhiro}, \au{Kobayashi, Hiromichi}, \au{Gotoda, Hiroshi} \& \au{Fukagata, Koji}} \yr{2025}  \at{Large-eddy simulation of high-{R}eynolds-number turbulent channel flow controlled using streamwise travelling wave-like wall deformation for drag reduction}.  \jt{Journal of Fluid Mechanics}  \bvol{1003},  \pg{A2}.

\bibitem[Nagib {\em et~al.\/}(2023)Nagib, Monkewitz \& Sreenivasan]{nagib2023reynolds}
{\sc \au{Nagib, Hassan}, \au{Monkewitz, Peter} \& \au{Sreenivasan, Katepalli~R}} \yr{2023}  \at{Reynolds number required to accurately discriminate between proposed trends of skin friction and normal stress in wall turbulence}.  \jt{arXiv preprint arXiv:2312.01184} .

\bibitem[Piomelli(1988)]{piomelli1988models}
{\sc \au{Piomelli, Ugo}} \yr{1988}  \at{Models for large eddy simulations of turbulent channel flows including transpiration}. PhD thesis, Stanford University.

\bibitem[Piomelli(2008)]{piomelli2008wall}
{\sc \au{Piomelli, Ugo}} \yr{2008}  \at{Wall-layer models for large-eddy simulations}.  \jt{Progress in Aerospace Sciences}  \bvol{44}~(6),  \pg{437--446}.

\bibitem[Piomelli \& Balaras(2002)]{piomelli2002wall}
{\sc \au{Piomelli, Ugo} \& \au{Balaras, Elias}} \yr{2002}  \at{Wall-layer models for large-eddy simulations}.  \jt{Annual Review of Fluid Mechanics}  \bvol{34}~(1),  \pg{349--374}.

\bibitem[Piomelli {\em et~al.\/}(2003)Piomelli, Balaras, Pasinato, Squires \& Spalart]{piomelli2003inner}
{\sc \au{Piomelli, Ugo}, \au{Balaras, Elias}, \au{Pasinato, Hugo}, \au{Squires, Kyle~D} \& \au{Spalart, Philippe~R}} \yr{2003}  \at{The inner--outer layer interface in large-eddy simulations with wall-layer models}.  \jt{International Journal of Heat and Fluid Flow}  \bvol{24}~(4),  \pg{538--550}.

\bibitem[Piomelli {\em et~al.\/}(1991)Piomelli, Cabot, Moin \& Lee]{piomelli1991subgrid}
{\sc \au{Piomelli, Ugo}, \au{Cabot, William~H}, \au{Moin, Parviz} \& \au{Lee, Sangsan}} \yr{1991}  \at{Subgrid-scale backscatter in turbulent and transitional flows}.  \jt{Physics of Fluids A: Fluid Dynamics}  \bvol{3}~(7),  \pg{1766--1771}.

\bibitem[Piomelli {\em et~al.\/}(1996)Piomelli, Yu \& Adrian]{piomelli1996subgrid}
{\sc \au{Piomelli, Ugo}, \au{Yu, Yunfang} \& \au{Adrian, Ronald~J}} \yr{1996}  \at{Subgrid-scale energy transfer and near-wall turbulence structure}.  \jt{Physics of Fluids}  \bvol{8}~(1),  \pg{215--224}.

\bibitem[Port{\'e}-Agel {\em et~al.\/}(2000)Port{\'e}-Agel, Meneveau \& Parlange]{porte2000scale}
{\sc \au{Port{\'e}-Agel, Fernando}, \au{Meneveau, Charles} \& \au{Parlange, Marc~B}} \yr{2000}  \at{A scale-dependent dynamic model for large-eddy simulation: application to a neutral atmospheric boundary layer}.  \jt{Journal of Fluid Mechanics}  \bvol{415},  \pg{261--284}.

\bibitem[Rezaeiravesh {\em et~al.\/}(2019)Rezaeiravesh, Mukha \& Liefvendahl]{rezaeiravesh2019systematic}
{\sc \au{Rezaeiravesh, Saleh}, \au{Mukha, Timofey} \& \au{Liefvendahl, Mattias}} \yr{2019}  \at{Systematic study of accuracy of wall-modeled large eddy simulation using uncertainty quantification techniques}.  \jt{Computers \& Fluids}  \bvol{185},  \pg{34--58}.

\bibitem[Schlatter \& {\"O}rl{\"u}(2010)]{schlatter2010assessment}
{\sc \au{Schlatter, Philipp} \& \au{{\"O}rl{\"u}, Ramis}} \yr{2010}  \at{Assessment of direct numerical simulation data of turbulent boundary layers}.  \jt{Journal of Fluid Mechanics}  \bvol{659},  \pg{116--126}.

\bibitem[Shankar {\em et~al.\/}(2025)Shankar, Chakraborty, Viswanathan \& Maulik]{shankar2025differentiable}
{\sc \au{Shankar, Varun}, \au{Chakraborty, Dibyajyoti}, \au{Viswanathan, Venkatasubramanian} \& \au{Maulik, Romit}} \yr{2025}  \at{Differentiable turbulence: Closure as a partial differential equation constrained optimization}.  \jt{Physical Review Fluids}  \bvol{10}~(2),  \pg{024605}.

\bibitem[Sirignano \& MacArt(2023)]{sirignano2023deep}
{\sc \au{Sirignano, Justin} \& \au{MacArt, Jonathan~F}} \yr{2023}  \at{Deep learning closure models for large-eddy simulation of flows around bluff bodies}.  \jt{Journal of Fluid Mechanics}  \bvol{966},  \pg{A26}.

\bibitem[Smagorinsky(1963)]{smagorinsky1963general}
{\sc \au{Smagorinsky, Joseph}} \yr{1963}  \at{General circulation experiments with the primitive equations: I. the basic experiment}.  \jt{Monthly Weather Review}  \bvol{91}~(3),  \pg{99--164}.

\bibitem[Smits {\em et~al.\/}(1983)Smits, Matheson \& Joubert]{smits1983low}
{\sc \au{Smits, AJ}, \au{Matheson, N} \& \au{Joubert, PN}} \yr{1983}  \at{Low-{R}eynolds-number turbulent boundary layers in zero and favorable pressure gradients}.  \jt{Journal of Ship Research}  \bvol{27}~(03),  \pg{147--157}.

\bibitem[Temmerman {\em et~al.\/}(2005)Temmerman, Had{\v{z}}iabdi{\'c}, Leschziner \& Hanjali{\'c}]{temmerman2005hybrid}
{\sc \au{Temmerman, L}, \au{Had{\v{z}}iabdi{\'c}, M}, \au{Leschziner, MA} \& \au{Hanjali{\'c}, K}} \yr{2005}  \at{A hybrid two-layer {URANS}--{LES} approach for large eddy simulation at high {R}eynolds numbers}.  \jt{International Journal of Heat and Fluid Flow}  \bvol{26}~(2),  \pg{173--190}.

\bibitem[Tessicini {\em et~al.\/}(2006)Tessicini, Temmerman \& Leschziner]{tessicini2006approximate}
{\sc \au{Tessicini, F}, \au{Temmerman, L} \& \au{Leschziner, MA}} \yr{2006}  \at{Approximate near-wall treatments based on zonal and hybrid {RANS}--{LES} methods for {LES} at high {R}eynolds numbers}.  \jt{International Journal of Heat and Fluid Flow}  \bvol{27}~(5),  \pg{789--799}.

\bibitem[Van~Driest(1956)]{van1956turbulent}
{\sc \au{Van~Driest, Edward~R}} \yr{1956}  \at{On turbulent flow near a wall}.  \jt{Journal of the Aeronautical Sciences}  \bvol{23}~(11),  \pg{1007--1011}.

\bibitem[Vreman(2004)]{vreman2004eddy}
{\sc \au{Vreman, AW}} \yr{2004}  \at{An eddy-viscosity subgrid-scale model for turbulent shear flow: Algebraic theory and applications}.  \jt{Physics of Fluids}  \bvol{16}~(10),  \pg{3670--3681}.

\bibitem[Wang {\em et~al.\/}(2017)Wang, Wu \& Xiao]{wang2017physics}
{\sc \au{Wang, Jian-Xun}, \au{Wu, Jin-Long} \& \au{Xiao, Heng}} \yr{2017}  \at{Physics-informed machine learning approach for reconstructing reynolds stress modeling discrepancies based on {DNS} data}.  \jt{Physical Review Fluids}  \bvol{2}~(3),  \pg{034603}.

\bibitem[Wang {\em et~al.\/}(2014)Wang, Hu \& Blonigan]{wang2014least}
{\sc \au{Wang, Qiqi}, \au{Hu, Rui} \& \au{Blonigan, Patrick}} \yr{2014}  \at{Least squares shadowing sensitivity analysis of chaotic limit cycle oscillations}.  \jt{Journal of Computational Physics}  \bvol{267},  \pg{210--224}.

\bibitem[Wang {\em et~al.\/}(2022)Wang, Wang \& Zaki]{wang2022observable}
{\sc \au{Wang, Qi}, \au{Wang, Mengze} \& \au{Zaki, Tamer~A}} \yr{2022}  \at{What is observable from wall data in turbulent channel flow?}  \jt{Journal of Fluid Mechanics}  \bvol{941},  \pg{A48}.

\bibitem[Wang {\em et~al.\/}(2021)Wang, Pan \& Wang]{wang2021energy}
{\sc \au{Wang, Wenkang}, \au{Pan, Chong} \& \au{Wang, Jinjun}} \yr{2021}  \at{Energy transfer structures associated with large-scale motions in a turbulent boundary layer}.  \jt{Journal of Fluid Mechanics}  \bvol{906},  \pg{A14}.

\bibitem[Weymouth \& Font(2025)]{WeymouthFont2025}
{\sc \au{Weymouth, G.D.} \& \au{Font, B.}} \yr{2025}  \at{Water{L}ily.jl: A differentiable and backend-agnostic {J}ulia solver for incompressible viscous flow around dynamic bodies}.  \jt{Computer Physics Communications}  \bvol{315},  \pg{109748}.

\bibitem[Whitmore {\em et~al.\/}(2020)Whitmore, Lozano-Dur{\'a}n \& Moin]{whitmore2020requirements}
{\sc \au{Whitmore, Michael}, \au{Lozano-Dur{\'a}n, Adri{\'a}n} \& \au{Moin, Parviz}} \yr{2020}  \at{Requirements and sensitivity analysis of {RANS}-free wall-modeled {LES}}.  \jt{arXiv preprint arXiv:2012.14900} .

\bibitem[Wu {\em et~al.\/}(2019)Wu, Xiao, Sun \& Wang]{wu2019reynolds}
{\sc \au{Wu, Jinlong}, \au{Xiao, Heng}, \au{Sun, Rui} \& \au{Wang, Qiqi}} \yr{2019}  \at{Reynolds-averaged {N}avier--{S}tokes equations with explicit data-driven reynolds stress closure can be ill-conditioned}.  \jt{Journal of Fluid Mechanics}  \bvol{869},  \pg{553--586}.

\bibitem[Yang {\em et~al.\/}(2019)Yang, Zafar, Wang \& Xiao]{yang2019predictive}
{\sc \au{Yang, XIA}, \au{Zafar, S}, \au{Wang, J-X} \& \au{Xiao, H}} \yr{2019}  \at{Predictive large-eddy-simulation wall modeling via physics-informed neural networks}.  \jt{Physical Review Fluids}  \bvol{4}~(3),  \pg{034602}.

\bibitem[Yang \& Griffin(2021)]{yang2021grid}
{\sc \au{Yang, Xiang~IA} \& \au{Griffin, Kevin~P}} \yr{2021}  \at{Grid-point and time-step requirements for direct numerical simulation and large-eddy simulation}.  \jt{Physics of Fluids}  \bvol{33}~(1).

\bibitem[Yang {\em et~al.\/}(2017)Yang, Park \& Moin]{yang2017log}
{\sc \au{Yang, Xiang~IA}, \au{Park, George~Ilhwan} \& \au{Moin, Parviz}} \yr{2017}  \at{Log-layer mismatch and modeling of the fluctuating wall stress in wall-modeled large-eddy simulations}.  \jt{Physical Review Fluids}  \bvol{2}~(10),  \pg{104601}.

\bibitem[Zhang {\em et~al.\/}(2026)Zhang, Yang \& He]{zhang2026differentiable}
{\sc \au{Zhang, Fengshun}, \au{Yang, Xiaolei} \& \au{He, Guowei}} \yr{2026}  \at{A differentiable wall-modeled large-eddy simulation method for high-{R}eynolds-number wall-bounded turbulent flows}.  \jt{Journal of Computational Physics}  \pg{p. 114835}.

\bibitem[Zhang {\em et~al.\/}(2022)Zhang, Xiao, Luo \& He]{zhang2022ensemble}
{\sc \au{Zhang, Xin-Lei}, \au{Xiao, Heng}, \au{Luo, Xiaodong} \& \au{He, Guowei}} \yr{2022}  \at{Ensemble {K}alman method for learning turbulence models from indirect observation data}.  \jt{Journal of Fluid Mechanics}  \bvol{949},  \pg{A26}.

\bibitem[Zhou \& Bae(2024)]{zhou2024sensitivity}
{\sc \au{Zhou, Di} \& \au{Bae, H~Jane}} \yr{2024}  \at{Sensitivity analysis of wall-modeled large-eddy simulation for separated turbulent flow}.  \jt{Journal of Computational Physics}  \bvol{506},  \pg{112948}.

\bibitem[Zhou \& Bae(2025)]{zhou2025effect}
{\sc \au{Zhou, Di} \& \au{Bae, H~Jane}} \yr{2025}  \at{Effect of subgrid-scale anisotropy on wall-modeled large-eddy simulation of separated flow}.  \jt{arXiv preprint arXiv:2511.18566} .

\end{thebibliography}


\end{document}